\documentclass[sn-basic]{sn-jnl}

\usepackage{graphicx}%
\usepackage{multirow}%
\usepackage{amsmath,amssymb,amsfonts}%
\usepackage{amsthm}%
\usepackage{mathrsfs}%
\usepackage[title]{appendix}%
\usepackage{xcolor}%
\usepackage{textcomp}%
\usepackage{manyfoot}%
\usepackage{booktabs}%
\usepackage{algorithm}%
\usepackage{algorithmicx}%
\usepackage{algpseudocode}%
\usepackage{listings}%
\usepackage{subcaption}
\usepackage{array}
\usepackage{booktabs} 

\DeclareUnicodeCharacter{202F}{ }

\theoremstyle{thmstyleone}%
\theoremstyle{thmstyletwo}%

\theoremstyle{thmstylethree}%

\begin{document}

\title[JOURNALDSP]{A Digital Twin of the FPGA Digital Signal Processing Chain for MKIDs Readout: Root-Cause Analysis and Mitigation of Spurs}


\author*[1]{\fnm{Mounir} \sur{Abdkrimi (corresponding author)}}\email{mounir.abdkrimi@gmail.com}

\author[2]{\fnm{Olivier} \sur{Rossetto}}

\author[2]{\fnm{Olivier} \sur{Bourrion}}

\author[2]{\fnm{Christophe} \sur{Vescovi}}

\author[2]{\fnm{Christophe} \sur{Hoarau}}

\affil*[1]{\orgdiv{Univ. Grenoble Alpes, CNRS, Grenoble INP, Institut Néel}, \city{Grenoble}, \postcode{38000}, \country{France}}
\affil[2]{\orgdiv{Univ. Grenoble Alpes, CNRS, LPSC-IN2P3}, \city{Grenoble}, \postcode{38000}, \country{France}}


\abstract{

The KID\_READOUT board, developed for the CONCERTO millimeter-wave astronomy instrument, implements FPGA-based digital frequency multiplexing to read out large arrays of Microwave Kinetic Inductance Detectors (MKIDs).
The complexity of the implemented multirate DSP chain, which combines tones synthesis, interpolation, digital frequency translation, polyphase filter-bank (PFB) channelization, and digital down-conversion (DDC), makes analytical performance optimization difficult.
To address this, we developed a cycle- and bit-accurate, Python-based digital twin of the FPGA readout firmware DSP chain.
Using this model, we identified the origin of previously measured and unexplained spurs in the readout channels, tracing them to periodicity mismatches between the excitation and analysis paths and to insufficient suppression of negative-frequency components by the DDC filters.
Based on these insights, we implemented a mitigation strategy that aligns the periodicities and improves the DDC filter characteristics, effectively eliminating the spurs with a minor increase in FPGA resource usage.

}



\maketitle

\section{Introduction}
\label{sec:Intro}

\subsection{Context and Motivation}

MKIDs have emerged as a promising technology for millimeter-wave detection~\cite{klutsch2003modelisation,Day2003}.
Over the past two decades, they have undergone substantial development, driven by the growing demand for highly sensitive observations in the millimeter-wave range (1–10\,mm).
This wavelength domain is crucial for exploring fundamental astrophysical processes, including the cosmic microwave background, star formation, and the evolution of galaxies~\cite{ward2007protostars}.

\begin{figure}[H]
    \centering    \includegraphics[width=1\textwidth]{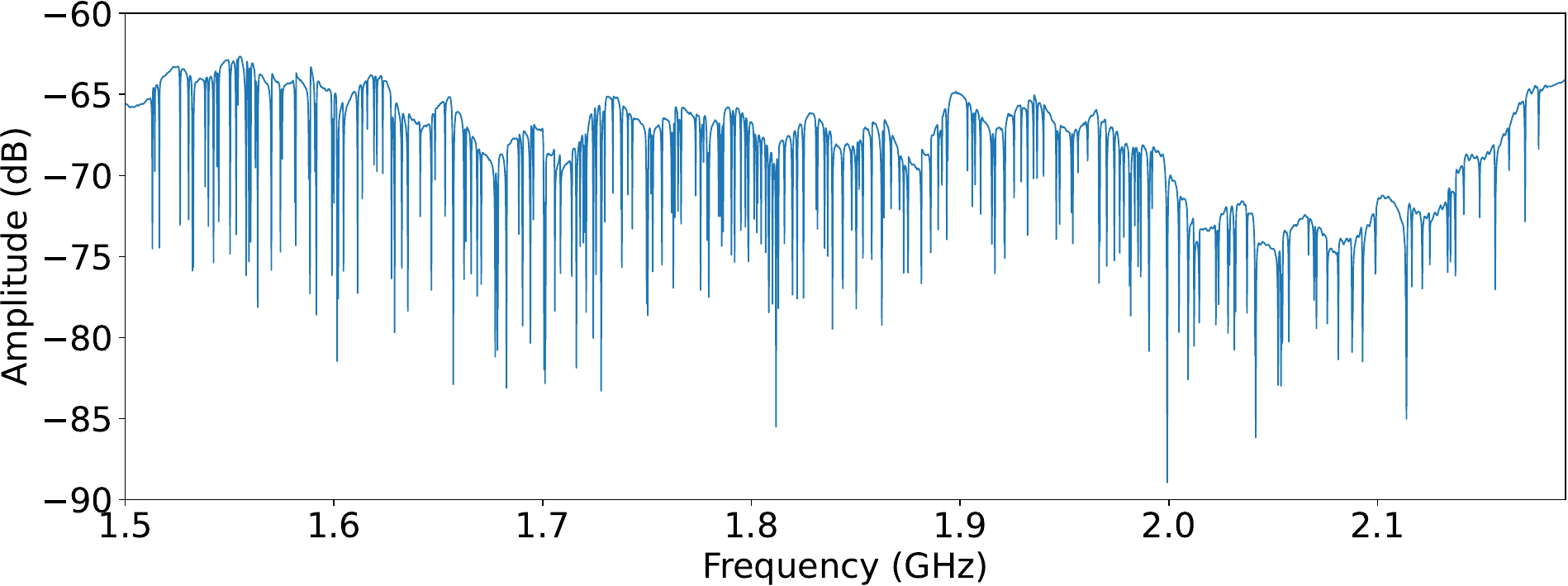}
    \caption{Frequency response of the MKID feedline used in the CONCERTO instrument.}
    \label{fig:freqres}
\end{figure}

These detectors are superconducting resonators composed of interdigitated capacitors and meandered inductors. 
When cooled below 100\,mK and operated in the radio-frequency (RF) domain, they exhibit high quality factors, enabling the multiplexing of thousands of resonators along a single transmission line. 
Each resonator functions as an independent pixel, uniquely identified by its resonance frequency.
Fig.~\ref{fig:freqres} shows the frequency response of CONCERTO’s MKIDs feedline, consisting of 400~resonators distributed across a 1\,GHz bandwidth. 

The instrument CONCERTO is a millimeter-wave, low spectral resolution imaging spectrometer with an instantaneous 18.6\,arcmin field of view and operates across the 130--310\,GHz frequency range~\cite{catalano2022concerto}. 
The instrument was deployed on the Atacama Pathfinder Experiment (APEX) telescope, where it collected science data between April~2021 and May~2023.

Next-generation instruments are expected to feature arrays with up to twice as many pixels, substantially increasing the demands on our readout electronics. 
To address these challenges, we developed a Python-based model of CONCERTO's KID\_READOUT DSP chain. 
This model enabled us to characterize current limitations and explore strategies for improvement.
Previous work discussed results obtained using this model~\cite{abdkrimi2025cordic, abdkrimi2025efficient,abdkrimi2024modeling}.

\subsection{Overview of MKID Readout Architecture}

\begin{figure}[H]
    \centering
    \includegraphics[width=0.9\textwidth,height=0.32\textwidth]{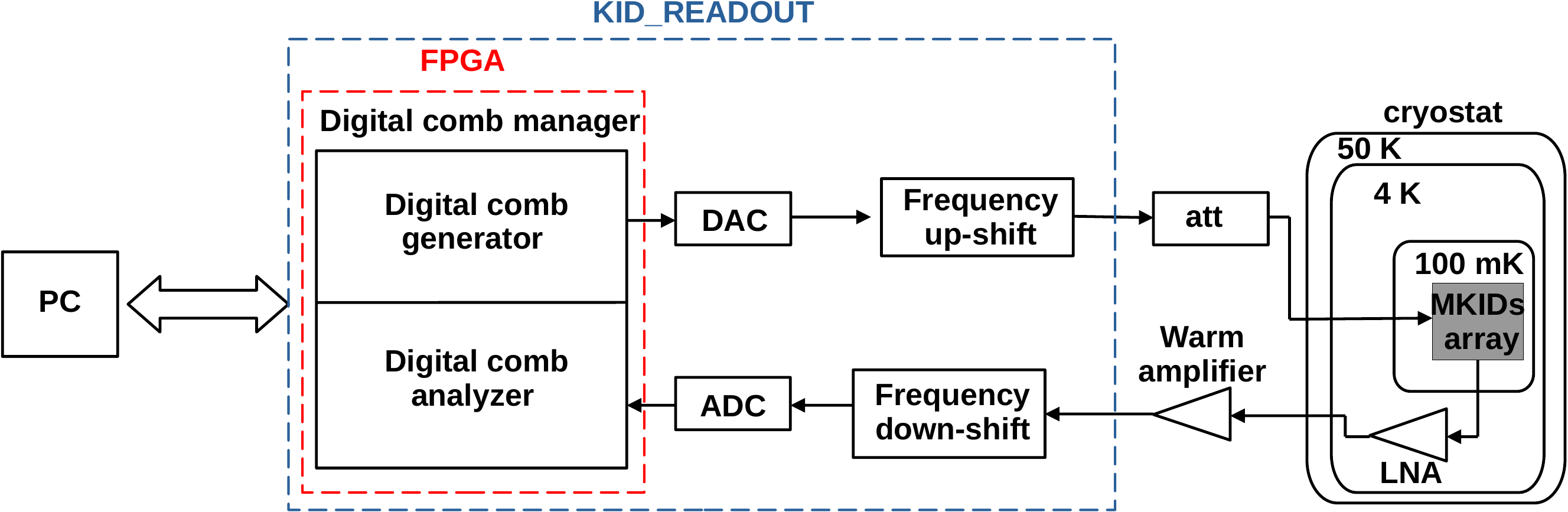}
    \caption{Overview of the instrumentation chain.}
    \label{fig: instrum}
\end{figure}

When a pixel absorbs a photon, the resonance frequency of its corresponding MKID shifts.
To measure the energy absorbed by each pixel in the array, the digital comb generator implemented within the FPGA (see Fig.~\ref{fig: instrum}) produces a digitally synthesized frequency comb composed of 400~excitation sinusoids, commonly referred to as tones in the MKID context.

Each tone is precisely tuned to the resonance frequency of an MKID in its dark (non-illuminated) state.
The digitally synthesized waveform is converted to analog, up-converted to the RF band of interest, and routed through attenuators at various temperature stages.
The signal returning from the MKIDs array is first amplified by a cryogenic low-noise amplifier; it is then further amplified at room temperature, down-converted to baseband, digitized by an analog-to-digital converter (ADC), and processed by the digital comb analyzer.
At this stage, the analyzer extracts the in-phase (I) and quadrature (Q) components of all 400~tones, from which the amplitude, $\sqrt{I^2 + Q^2}$, and phase, $\arctan\left(\frac{Q}{I}\right)$, are computed offline on a PC.
These two quantities provide a measure of the energy absorbed by the corresponding MKID.
Further technical details on the KID\_READOUT board architecture and performance are available in previous studies~\cite{bourrion2022concerto,bounmy2022concerto}.

\subsection{Contribution of this work}

In this paper, we present a Python-based digital twin of the digital comb manager that faithfully reproduces both the functional behavior and the architectural details of the existing firmware, providing a bit-accurate and cycle-accurate model of the DSP chain implemented on the FPGA.

The main motivation behind this approach is that the Python-based simulation generates sufficient data for our entire DSP chain in approximately one hour, whereas the equivalent VHDL simulation requires tens of hours. 
This makes it possible to design fully parameterizable DSP blocks and to run extensive simulations in order to analyze the DSP chain performance under different block configurations.
In addition, Python offers convenient built-in tools for data visualization and performance analysis within the same environment.

Thanks to the high fidelity of the model, the digital twin accurately reproduces the spurious tones observed in actual CONCERTO instrument data. This capability enabled a systematic analysis of the artifacts within the DSP chain, revealing two main sources: periodicity mismatches between the excitation and analysis paths, and insufficient suppression of negative-frequency components in the digital downconversion stage.
Leveraging these insights, we designed and evaluated—entirely within the Python model—a mitigation strategy consisting of (i) aligning the periodicities of the excitation and analysis paths and (ii) redesigning the DDC filters.
The proposed modifications were subsequently implemented on the KID\_READOUT's FPGA and validated through measurements, resulting in a significant improvement in spectral purity.

\section{Digital Twin: Simulation, Analysis, and Mitigation of Spurs}
\label{sec:consolidated_digital}

\subsection{Simulation}
\label{subsec:consolidated_chain_sim}

This section presents simulation results obtained with the digital twin. The goal is to assess the end-to-end performance of the complete digital processing chain under realistic operating conditions. 
To this end, the model is operated in a closed-loop configuration, where the output of the excitation path is directly fed back into the input of the analysis path.

The simulation replicates normal CONCERTO operation.
In the excitation chain, the model constructs the excitation waveform following the same hierarchical structure as the FPGA implementation. Ten frequency subbands, each spanning 100\,MHz of bandwidth, are first generated in parallel, with each subband containing 40~tones. Each tone is individually produced by a modeled CORDIC block, and the tones within a given subband are summed to form a subband waveform.
The resulting ten parallel subband waveforms are then interpolated and numerically frequency-shifted to position the ten subbands at their intended spectral locations. Finally, all subbands are combined to form a single wideband digital waveform spanning 1\,GHz and comprising 400~tones.

In the analysis chain, the looped-back excitation waveform is processed by a modeled polyphase filter bank, which channelizes it into ten subbands. Each subband is then routed to 40 dedicated modeled DDCs. For each tone, the corresponding DDC reuses the reference tone generated by the matching CORDIC block in the excitation chain to perform numerical down-conversion to baseband (0\,Hz).

\begin{figure}[h]
    \centering
    \begin{subfigure}[t]{0.98\textwidth}
        \centering
        \begin{subfigure}[t]{0.49\textwidth}
            \centering
            \includegraphics[width=\textwidth]{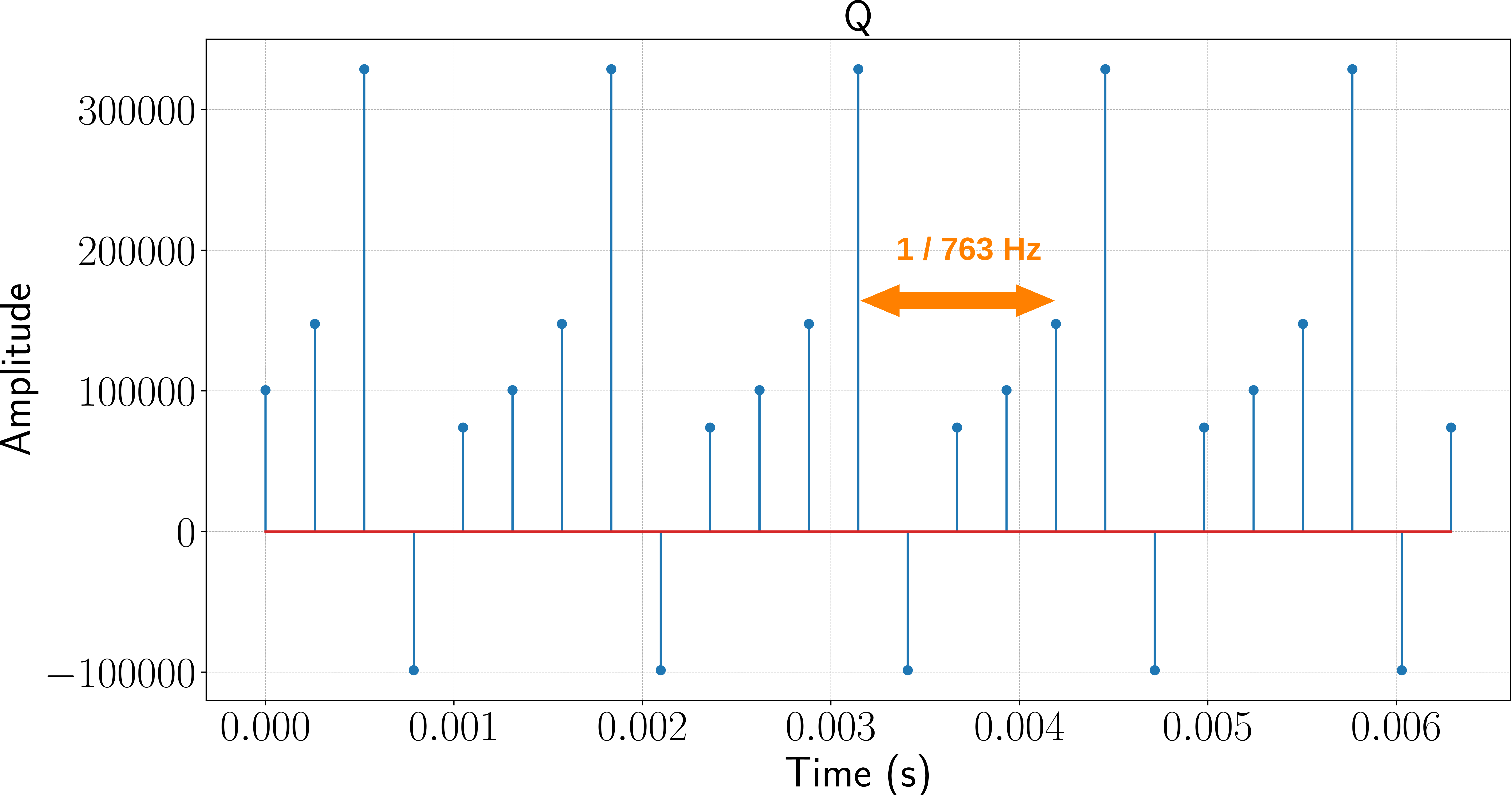}
        \end{subfigure}
        \hfill
        \begin{subfigure}[t]{0.49\textwidth}
            \centering
            \includegraphics[width=\textwidth]{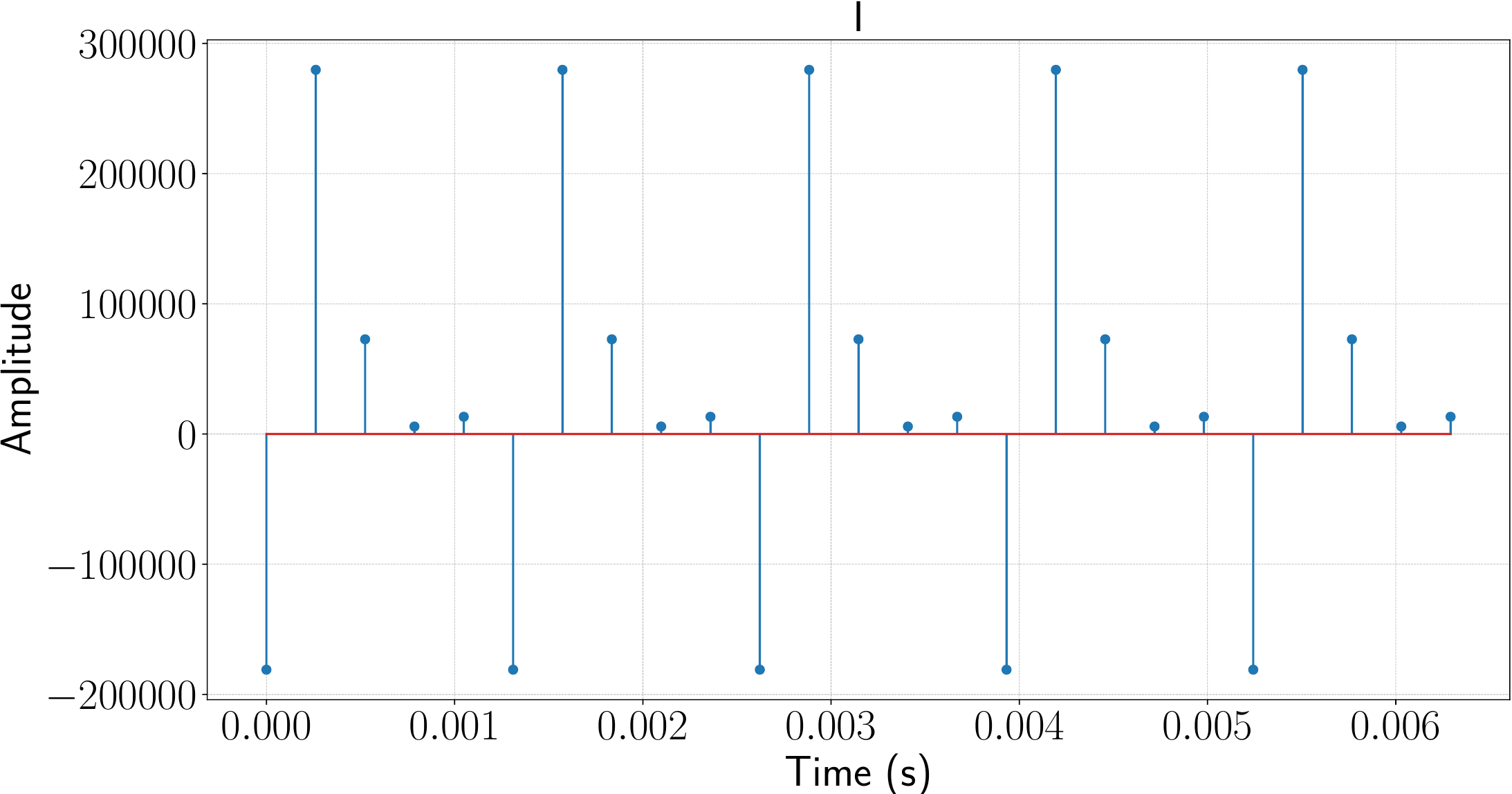}
        \end{subfigure}
        \caption{}
        \label{fig:setup1a}
    \end{subfigure}

    \vspace{0.3cm}

    \begin{subfigure}[t]{0.98\textwidth}
        \centering
        \begin{subfigure}[t]{0.49\textwidth}
            \centering
            \includegraphics[width=\textwidth]{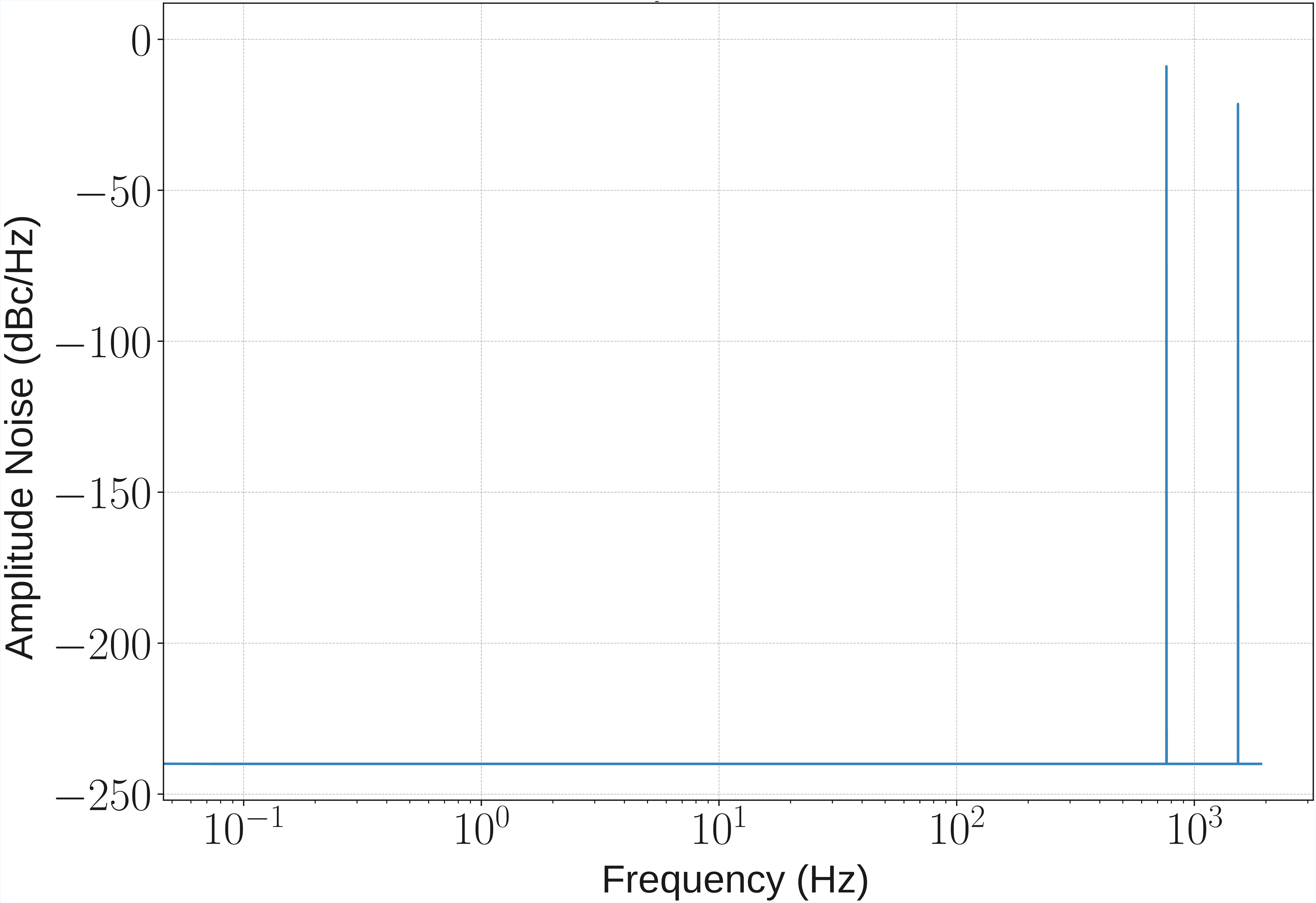}
        \end{subfigure}
        \hfill
        \begin{subfigure}[t]{0.49\textwidth}
            \centering
            \includegraphics[width=\textwidth]{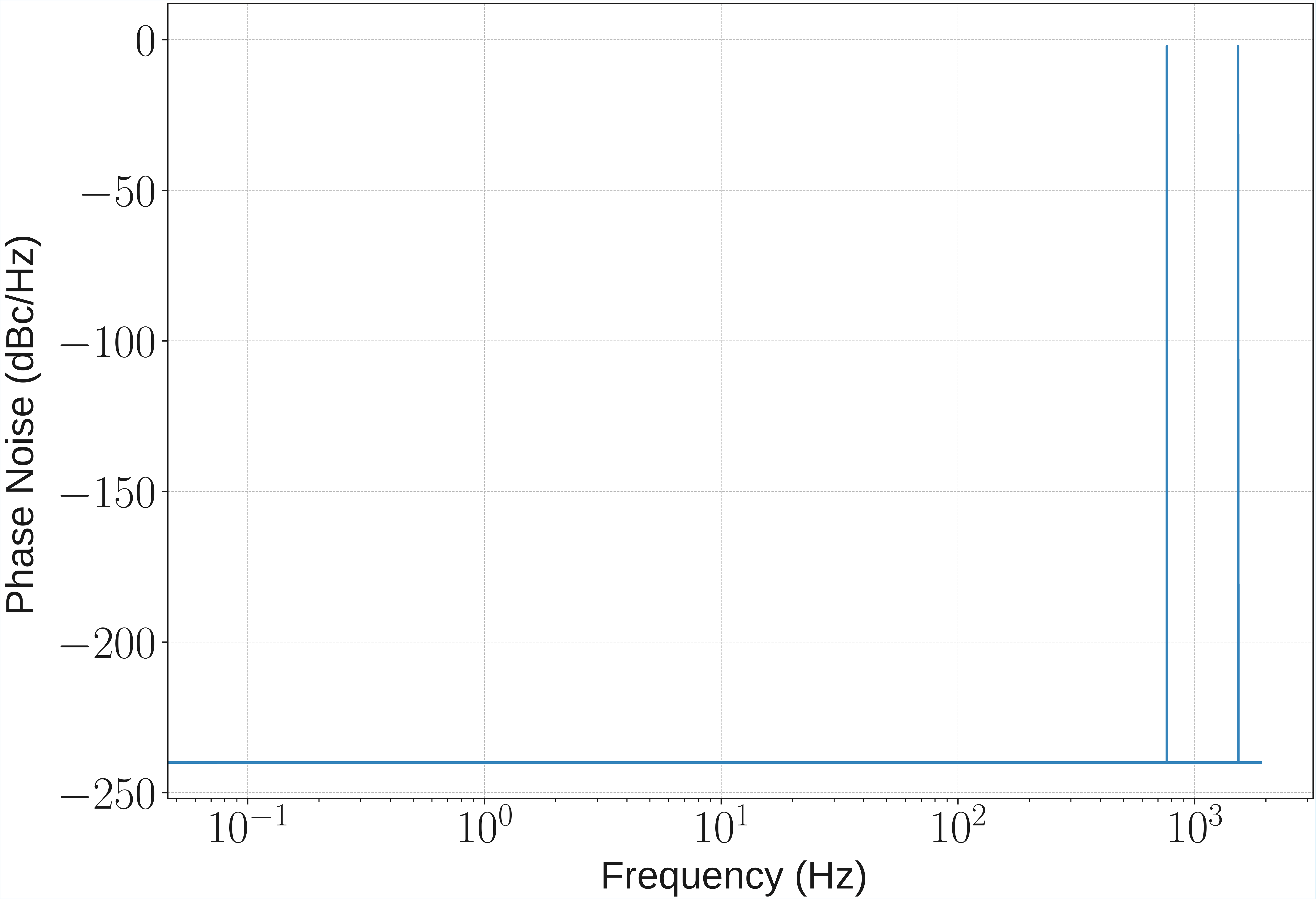}
        \end{subfigure}
        \caption{}
        \label{fig:setup1b}
    \end{subfigure}

\caption{Simulation results: (a) Time-domain representation of one selected I/Q signal; (b) corresponding amplitude and phase PSD.}

    \label{fig:simu_resu_digital}
\end{figure}

The resulting 400~complex I/Q data streams, produced by the DDCs and sampled at $f_s = 3814\,\text{Hz}$, are subsequently analyzed to evaluate signal integrity and to identify any artifacts introduced by the digital processing chain.
Fig.~\ref{fig:simu_resu_digital}(a) shows the time-domain I/Q signals for a representative tone.
Since each DDC demodulates its assigned tone to 0\,Hz and applies low-pass filtering, the ideal I/Q outputs are expected to be constant. 
Instead, a periodic fluctuation with a five-sample cycle is consistently observed across all 400~I/Q data streams.
Fig.~\ref{fig:simu_resu_digital}(b) shows the corresponding power spectral densities (PSD) of the amplitude and phase noise derived from these I/Q signals.

The amplitude and phase noise are obtained by, first, converting the I/Q data into polar form: the amplitude is calculated as \(\sqrt{I^2 + Q^2}\), and the phase as \(\arctan(Q/I)\).
Then, to isolate the noise components, the mean amplitude is subtracted from the raw amplitude, and the resulting signal is normalized. The same procedure is applied to the phase.

Two prominent spurious components are observed at approximately 763\,Hz and 1526\,Hz, corresponding more precisely to 762.94\,Hz and 1525.88\,Hz, respectively. 
The 763\,Hz component corresponds to the five-sample periodicity observed in the time domain, given by \(f_s / 5 \approx 3814\,\text{Hz} / 5 \approx 763\,\text{Hz}\). The 1526\,Hz component is identified as its second harmonic.

\begin{figure}[h]
    \centering
    \begin{subfigure}[t]{0.49\textwidth}
        \centering
        \includegraphics[width=\textwidth]{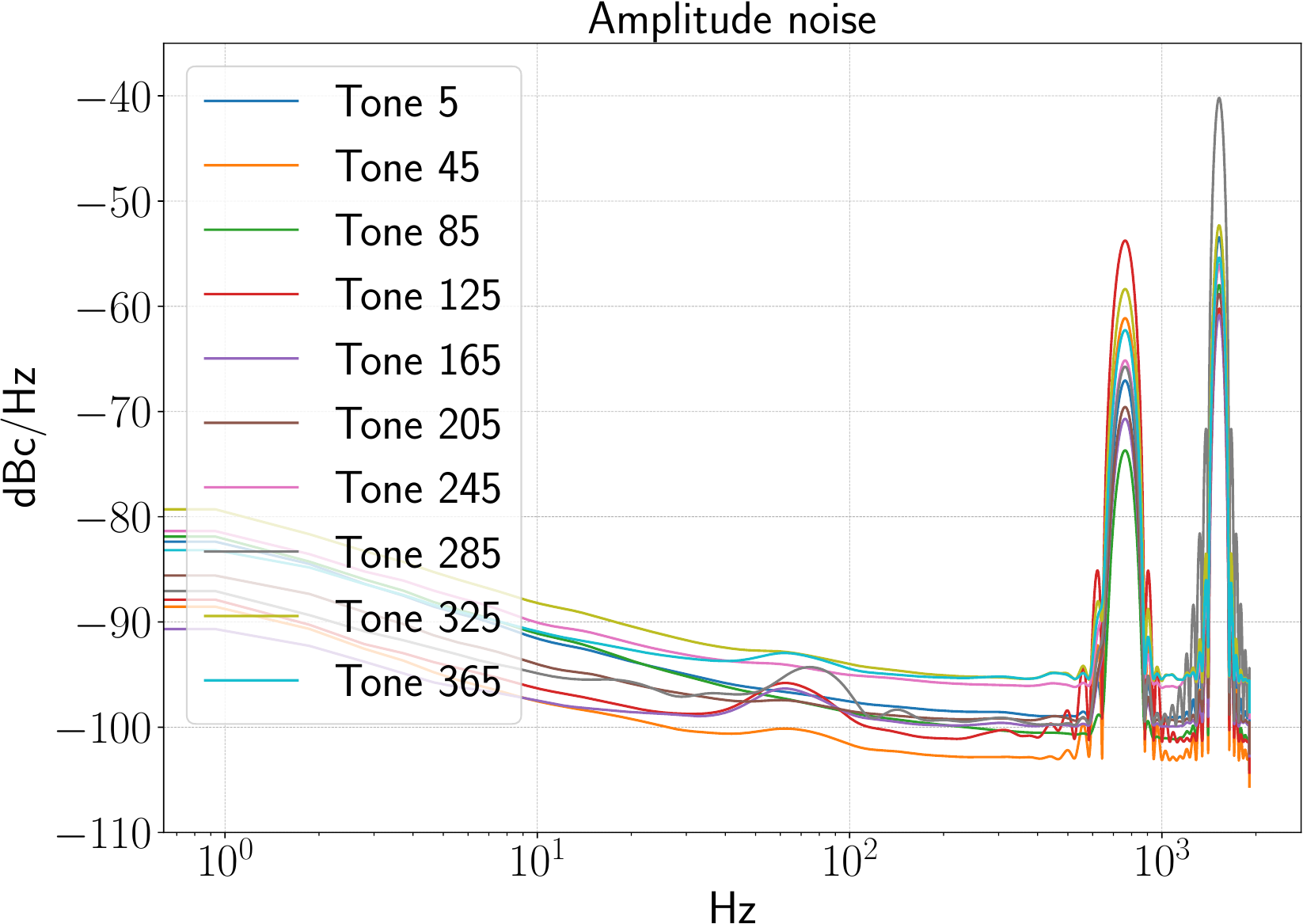}
    \end{subfigure}
    \hfill
    \begin{subfigure}[t]{0.49\textwidth}
        \centering
        \includegraphics[width=\textwidth]{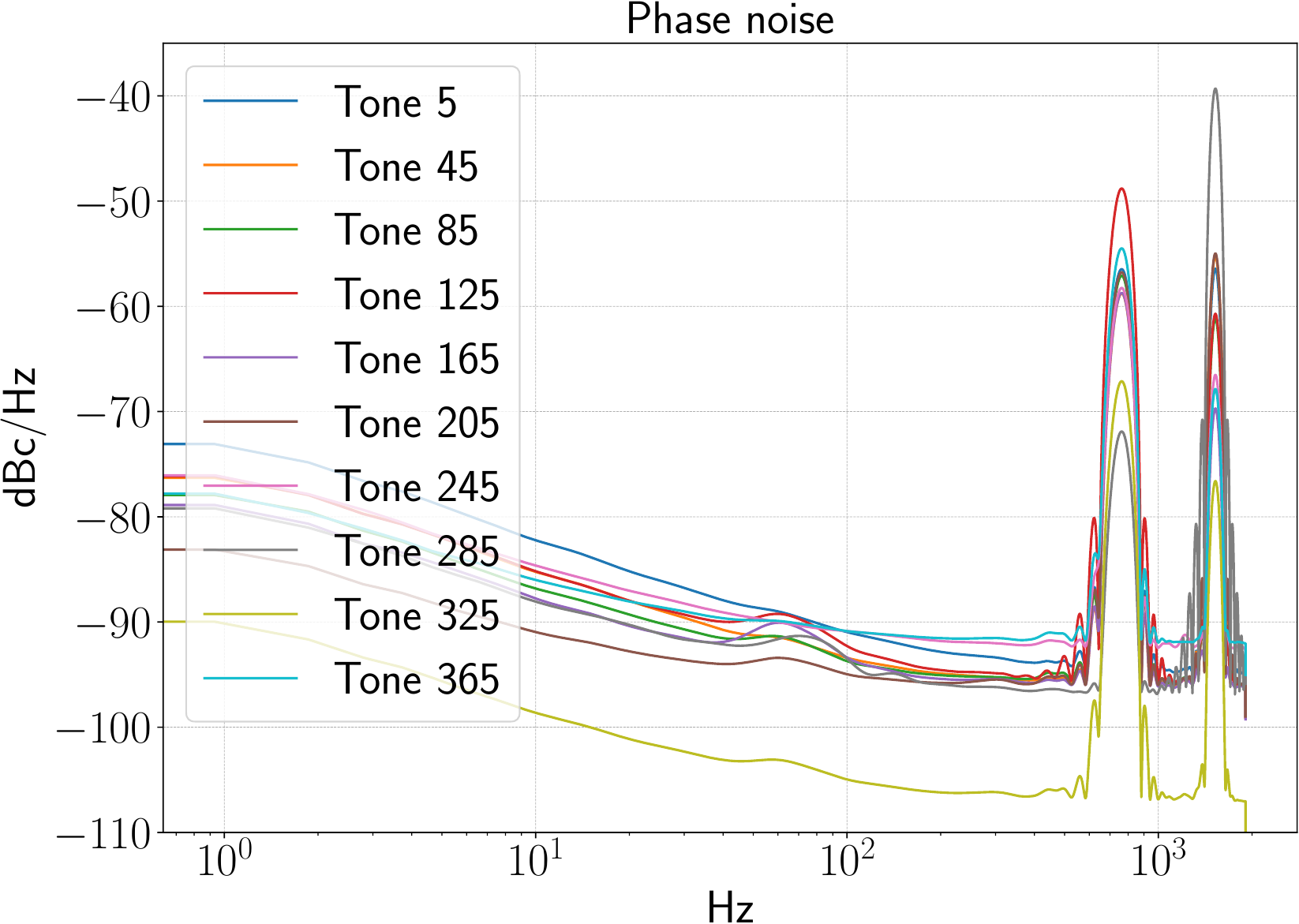}
    \end{subfigure}
    \caption{PSD measurement of amplitude and phase noise from CONCERTO~\cite{bourrion2022concerto}. 
    A representative tone from each of the 10~frequency bands is shown.}
    \label{fig:phenoCONCERTO}
\end{figure}

Crucially, the same spurious components have also been observed in CONCERTO measurement data, as illustrated in Fig.~\ref{fig:phenoCONCERTO}, and were previously reported~\cite{bourrion2022concerto}; however, their origin remained unclear.

The strong agreement between simulation and measurement not only validates the accuracy of the twin model but also provides compelling evidence that these spurious tones originate from within the digital system itself.
This finding has prompted a deeper investigation into the root causes of the observed spurs and the development of an appropriate mitigation strategy, as detailed in Section~\ref{subsec:phenomenon_explanation}.

\subsection{Origin of Spurs Analysis} \label{subsec:phenomenon_explanation}

To identify the root cause of these artifacts, we analyze the signal behavior at each stage of the digital processing chain using the digital twin, in both the time and frequency domains.
This investigation is conducted using the same closed-loop configuration as in Section~\ref{subsec:consolidated_chain_sim}. 
However, for clarity, two simplifications are introduced.
First, the simulation is reduced to a single-tone scenario instead of the full configuration involving ten subbands and 400 tones.
Second, the floating-point version of the model is used in order to obtain cleaner spectra at the output of each processing stage.
These simplifications do not alter the underlying spectral phenomenon under investigation. Rather, they enable a clearer isolation and interpretation of the mechanisms responsible for the observed spurious tones.

\subsubsection{Tone Generator}

The excitation chain begins with the tone generator, consisting of a 16-bit phase accumulator and a CORDIC unit that converts the accumulated phase into (I/Q) amplitude outputs.

\paragraph{Time Domain:}

The phase accumulator is incremented by a Frequency Control Word (FCW) on each clock cycle, with the FPGA on KID\_READOUT operating at a sampling frequency of 250\,MHz.
When the accumulator exceeds its maximum value of \(2^{16} - 1\), it wraps around to zero, plus the excess.  
This behavior renders the output of the phase accumulator—and thus the tone generator—periodic, with a maximum period of \(2^{16}\) samples, as the accumulator cycles through the same sequence of phase values repeatedly.
 
\begin{figure}[H]
    \centering
    \includegraphics[width=1\textwidth]{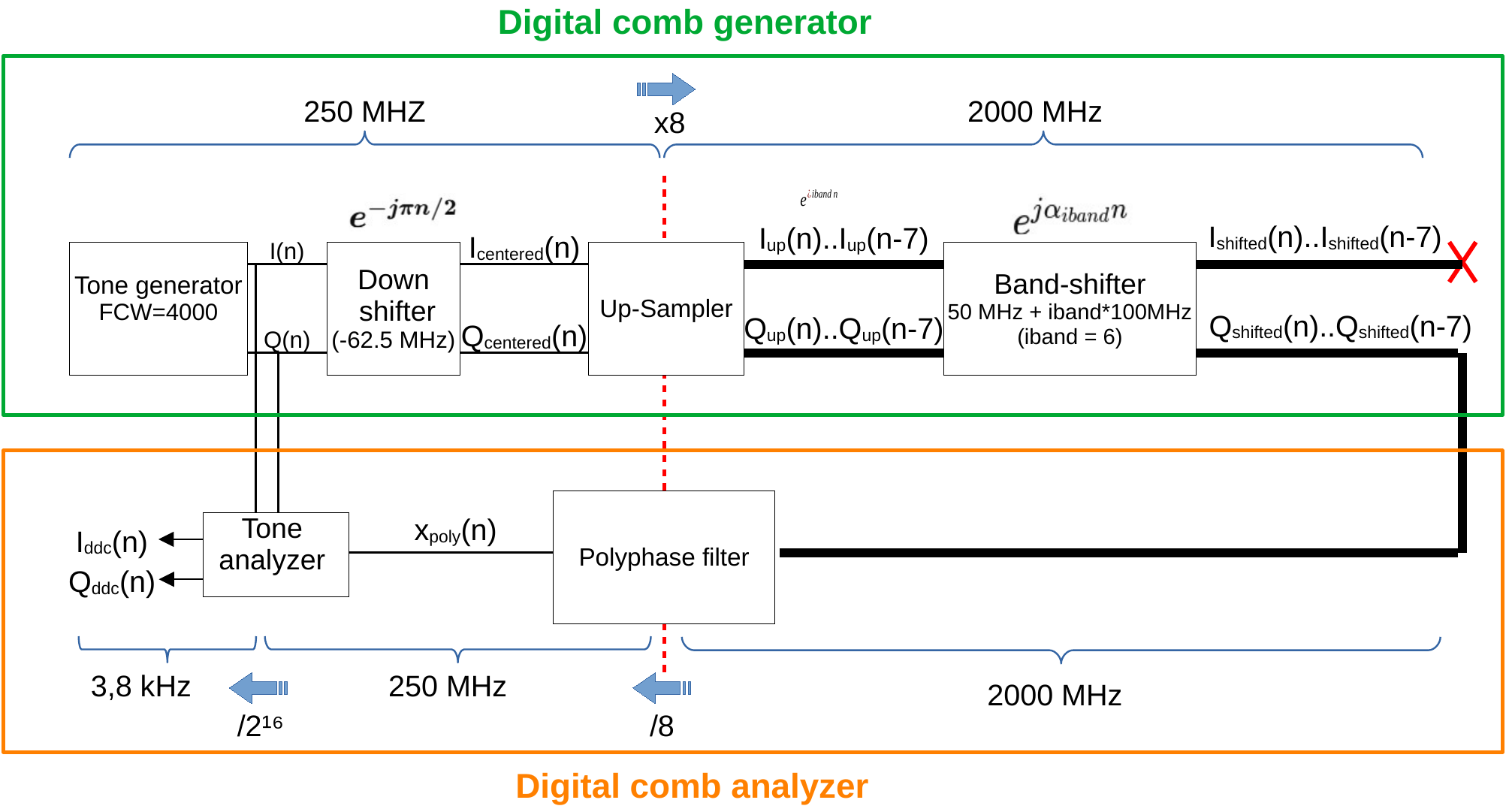}\caption{Digital loop-back : Block-level description of the simulated digital excitation and analysis chains operated in closed loop for a single-tone configuration.}

    \label{fig:blocks_anal}
\end{figure}

In the single-tone configuration considered in this study, the FCW is arbitrarily chosen to be 4000 as illustrated in Fig.~\ref{fig:blocks_anal}. This choice corresponds to a tone frequency of :
\[
f_1 = \frac{250 \text{ MHz}}{2^{16}} \times 4000 \approx 15.26\text{ MHz}.
\]

The periodic nature of the generated signal is verified by comparing multiple consecutive segments of the signal, which confirms that
\[
I[n] = I[n + 2^{16}], \quad Q[n] = Q[n + 2^{16}] \quad \forall n
\]

\paragraph{Frequency Domain:}

The spectral components of the signal (either $I$ or $Q$) align exactly with the bins of the discrete Fourier transform when a $2^{16}$-point FFT is applied.
To avoid redundancy, we show only one of the spectra—either $I$ or $Q$—throughout this section, as their amplitude spectra are identical.

\begin{figure}[H]
    \centering
    \begin{subfigure}[t]{0.49\textwidth}
        \centering
        \includegraphics[width=\textwidth]{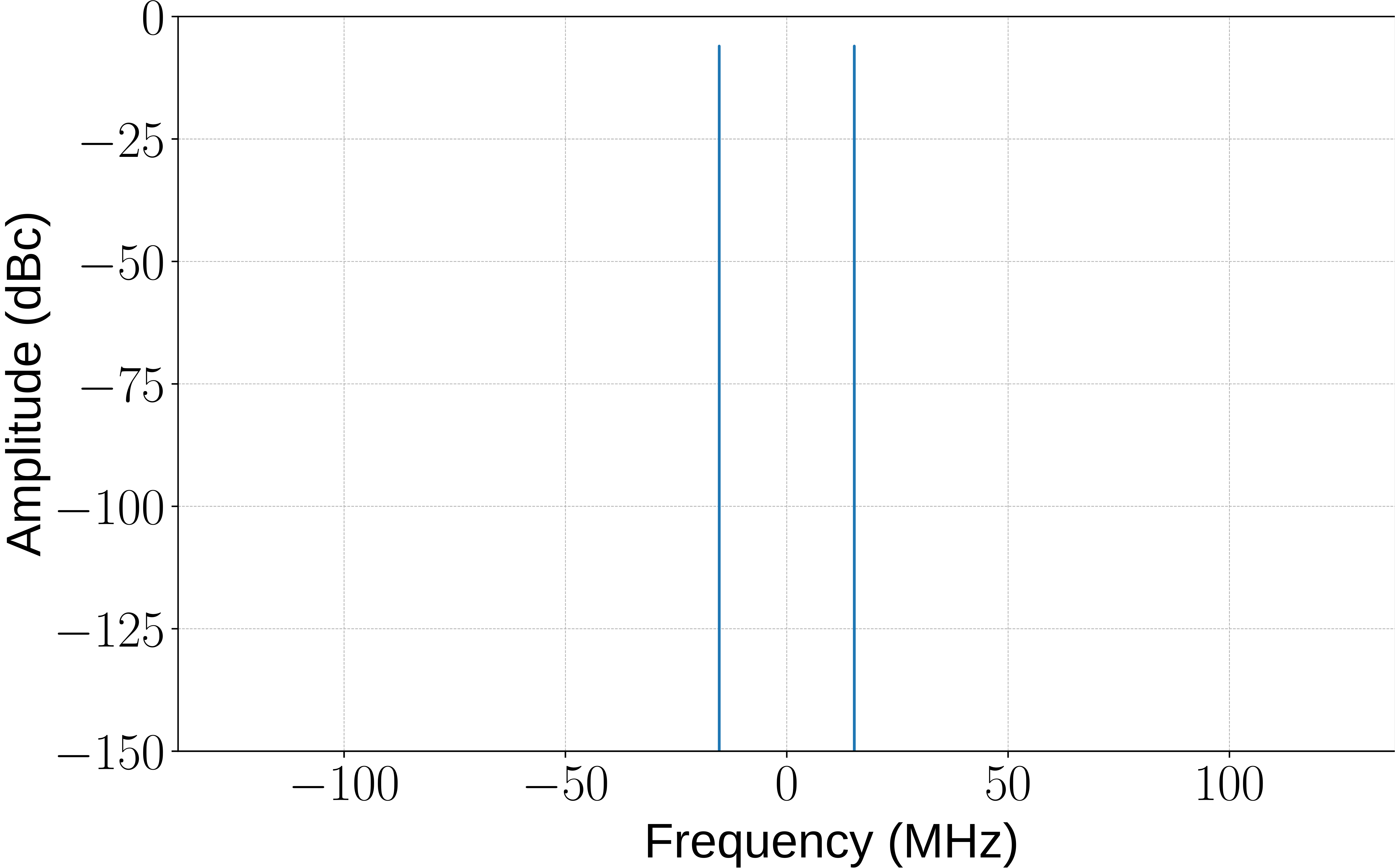}
        \caption{FFT with \(2^{16}\) samples: sharp bin alignment.}
        \label{fig:pheno1a}
    \end{subfigure}
    \hfill
    \begin{subfigure}[t]{0.49\textwidth}
        \centering
        \includegraphics[width=\textwidth]{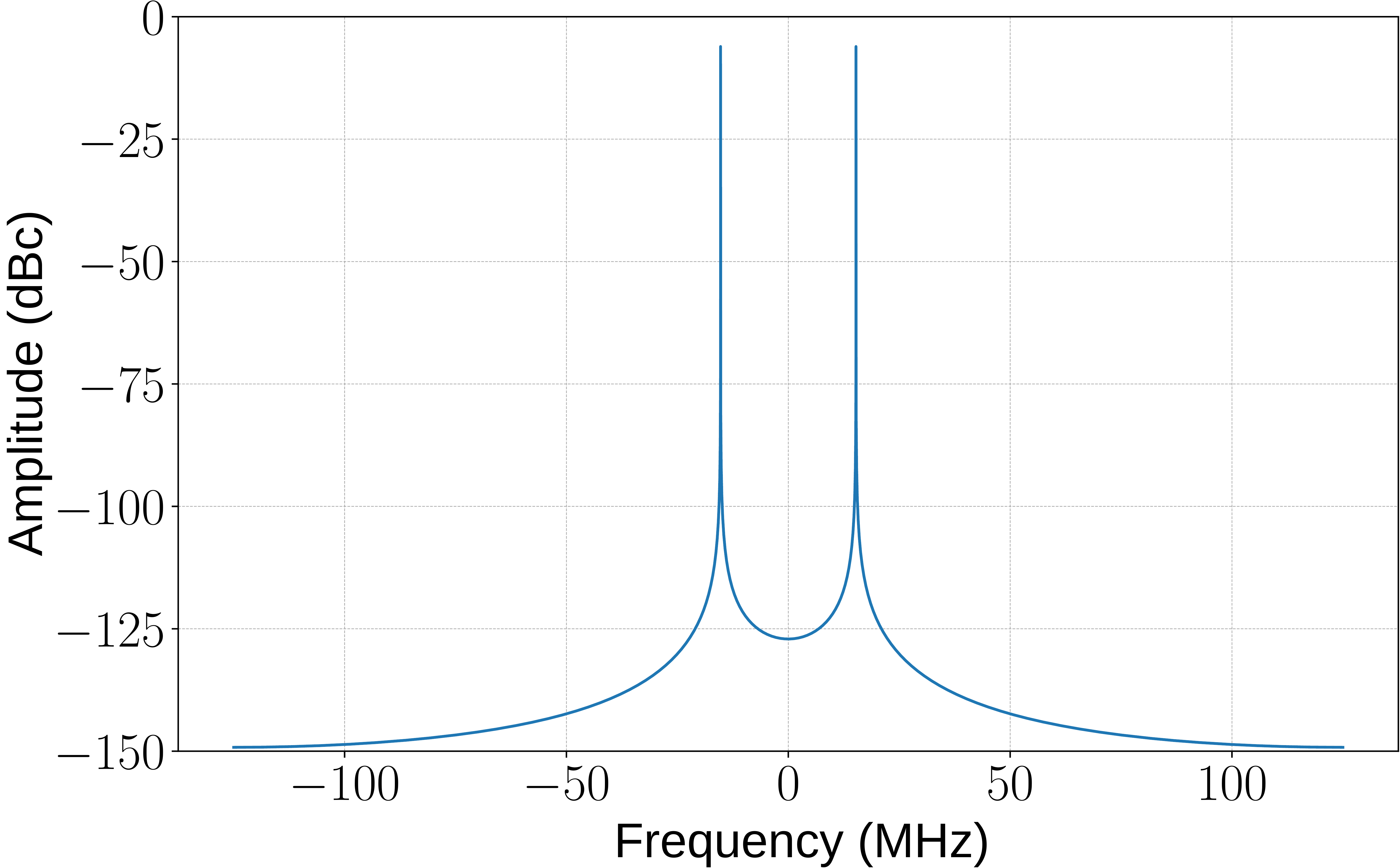}
        \caption{FFT with \(2^{16} - 1\) samples: spectral broadening.}
        \label{fig:pheno1b}
    \end{subfigure}
    \caption{FFT spectra of the tone generator output with sampling rate \(F_s = 250\,\text{MHz}\).}
    \label{fig:pheno1}
\end{figure}

This implies that the tone frequency is an integer multiple of the FFT’s frequency resolution, which is given by $250\,\text{MHz} / 2^{16}$.  
Fig.~\ref{fig:pheno1}(a) illustrates this bin alignment, showing two distinct spectral lines at $\pm15.26\,\text{MHz}$.
In contrast, Fig.~\ref{fig:pheno1}(b) shows the spectral broadening that occurs when the FFT is computed over $2^{16} - 1$ samples.
Although this broadening example may not appear important to mention, it will be used throughout the analysis as a means to reveal changes in signal periodicity and will play a key role in later interpretations.

\subsubsection{Down-shifter}

The tone generator produces I/Q signals within the frequency range of \([12.5\,\text{MHz},\,112.5\,\text{MHz}]\). 
A digital frequency shift operation is then applied to relocate this range symmetrically around zero—i.e., to $[-50\,MHz-50\,MHz]$.
This operation is carried out by the down-shifter illustrated in Fig.~\ref{fig:blocks_anal}, which translates the complex signal
\[
x[n] = I[n] + jQ[n]
\]
by \(-62.5\,\text{MHz}\) through complex multiplication with the phasor
\[
e^{-j\frac{\pi}{2}n},
\]
which corresponds to a frequency shift of \(-\frac{F_s}{4} = -62.5\,\text{MHz}\) for a sampling rate of \(F_s = 250\,\text{MHz}\).
The resulting centered signal is thus given by
\[
x_{\text{centered}}[n] = x[n] \cdot e^{-j\frac{\pi}{2}n}.
\]

Expanding this multiplication in terms of real and imaginary components:

$$
\begin{aligned}
x_{\text{centered}}[n] &= \left[I[n] + jQ[n]\right] \cdot \left[\cos\left(\tfrac{\pi}{2}n\right) - j\sin\left(\tfrac{\pi}{2}n\right)\right] \\
&= \big(I[n] \cos\left(\tfrac{\pi}{2}n\right) + Q[n] \sin\left(\tfrac{\pi}{2}n\right)\big) + j\big(Q[n] \cos\left(\tfrac{\pi}{2}n\right) - I[n] \sin\left(\tfrac{\pi}{2}n\right)\big).
\end{aligned}
$$

Thus, the new real in-phase and quadrature components become:

$$
\begin{aligned}
I_{\text{centered}}[n] &= I[n] \cos\left(\tfrac{\pi}{2}n\right) + Q[n] \sin\left(\tfrac{\pi}{2}n\right), \\
Q_{\text{centered}}[n] &= Q[n] \cos\left(\tfrac{\pi}{2}n\right) - I[n] \sin\left(\tfrac{\pi}{2}n\right).
\end{aligned}
$$

\paragraph{Time Domain:}

This operation does not alter the signal $2^{16}$ sample periodicity.
As discussed, the tone generator output is periodic with a period of $2^{16}$ samples, while the modulation phasor $e^{-j\frac{\pi}{2}n}$ has a period of 4~samples.
The resulting signal after modulation has a periodicity equal to the least common multiple (LCM) of these two values:

$$
\text{LCM}(2^{16}, 4) = 2^{16}.
$$

Since 4~divides evenly into $2^{16}$, the overall periodicity remains $2^{16}$ samples.
Thus, the frequency shift operation preserves the original time-domain periodicity.
Consecutive \(2^{16}\)-sample segments of the in-phase and quadrature signals were analyzed, confirming this periodicity:

\[
I_{\text{centered}}[n] = I_{\text{centered}}[n + 2^{16}], \quad Q_{\text{centered}}[n] = Q_{\text{centered}}[n + 2^{16}] \quad \forall n
\]

\paragraph{Frequency Domain:} 

In the specific case of the tone at $f_1 = 15.26\,\text{MHz}$, this complex modulation results in a new frequency of :

$$
f_{\text{shifted}} = f_1 - 62.5\,\text{MHz} = -47.24\,\text{MHz}.
$$

\begin{figure}[h]
    \centering
    \begin{subfigure}[t]{0.49\textwidth}
        \centering
        \includegraphics[width=\textwidth]{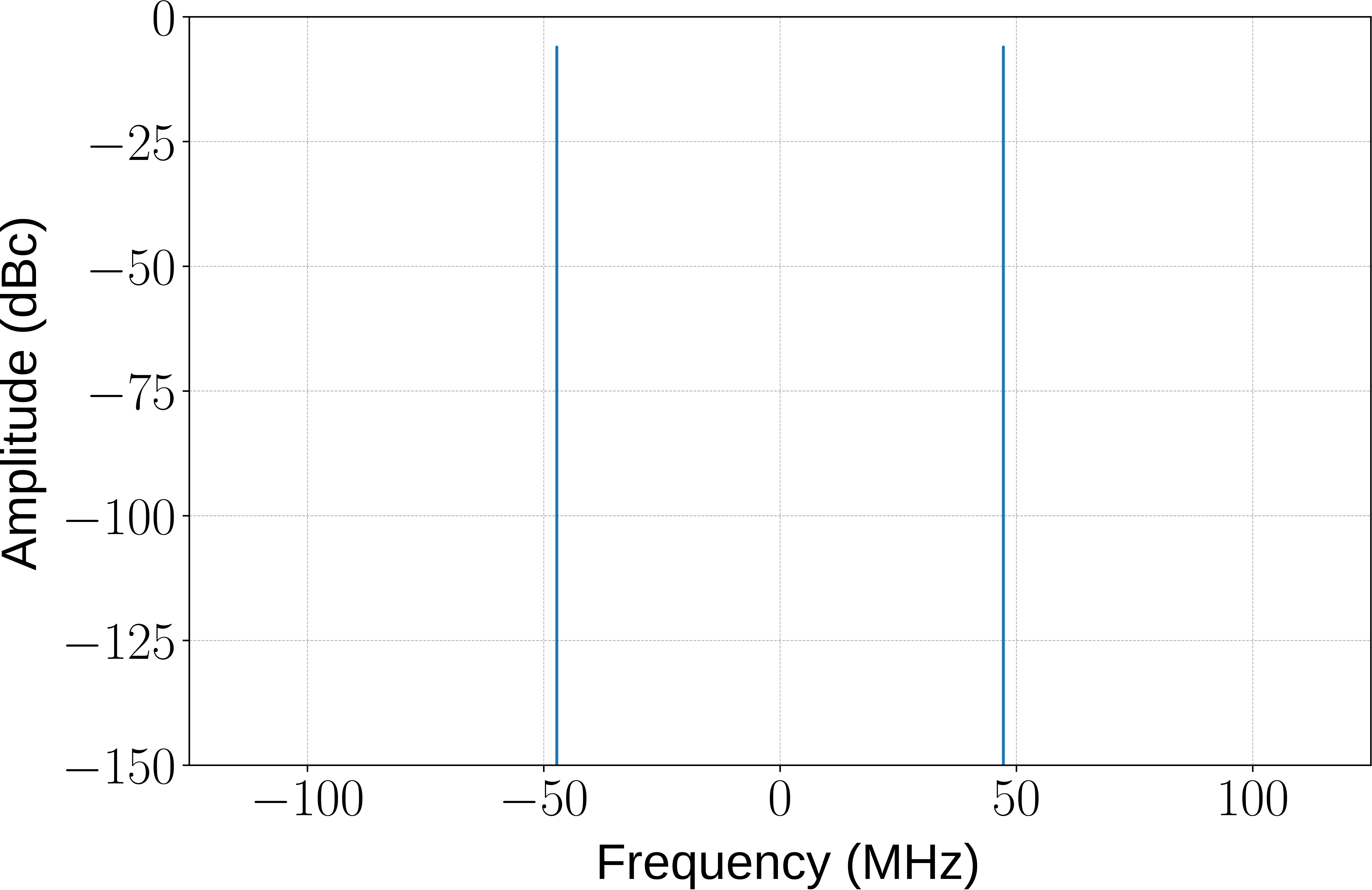}
        \caption{FFT with \(2^{16}\) samples: sharp bin alignment.}
        \label{fig:pheno200a}
    \end{subfigure}
    \hfill
    \begin{subfigure}[t]{0.49\textwidth}
        \centering
        \includegraphics[width=\textwidth]{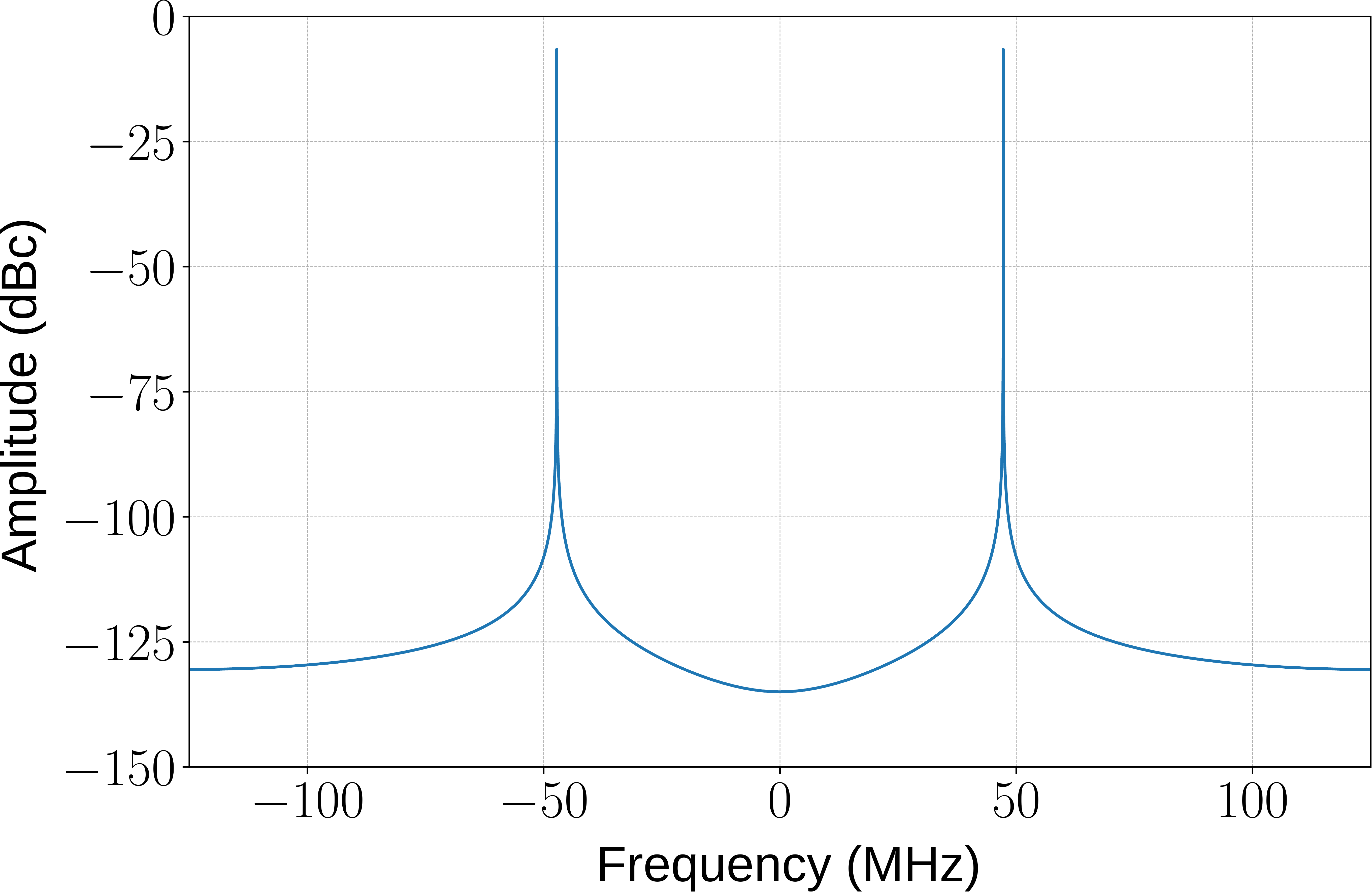}
        \caption{FFT with \(2^{16} - 1\) samples: spectral broadening.}
        \label{fig:pheno200b}
    \end{subfigure}
    \caption{FFT spectra of the frequency-centered tone at $-47.24\,\text{MHz}$ with a sampling rate of $F_s = 250\,\text{MHz}$.
 }
    \label{fig:pheno200}
\end{figure}

Because $I[n]$ and $Q[n]$ are real-valued signals, their spectra exhibit conjugate symmetry.
Fig.~\ref{fig:pheno200} shows the spectral outcome after frequency centering, where two symmetrical peaks appear at $\pm47.24\,\text{MHz}$, consistent with the expected frequency shift.

The figure also demonstrates that using an FFT size of $2^{16}$ results in perfect bin alignment, confirming that the signal’s periodicity is preserved. 
In contrast, applying an FFT with $2^{16} - 1$ points leads to spectral broadening.

\subsubsection{Upsampling: Periodicity extension}

The centered in-phase and quadrature signals are upsampled by a factor of~8, increasing the sampling rate from 250\,MHz to 2000\,MHz.

\paragraph{Time Domain:}

The upsampler processes the same \(2^{16}\)-sample sequence repeatedly while applying identical interpolation operations; consequently, the resulting upsampled signal exhibits a new periodicity of \(8 \times 2^{16}\) samples.
We compared multiple consecutive \(8 \times 2^{16}\) segments, and confirmed the extended periodicity :

\[
I_{\text{up}}[n] = I_{\text{up}}[n + 8 \times 2^{16}], \quad 
Q_{\text{up}}[n] = Q_{\text{up}}[n + 8 \times 2^{16}] \quad \forall n
\]

\paragraph{Frequency Domain:} 

\begin{figure}[h]
    \centering
    \begin{subfigure}[t]{0.49\textwidth}
        \centering
        \includegraphics[width=\textwidth]{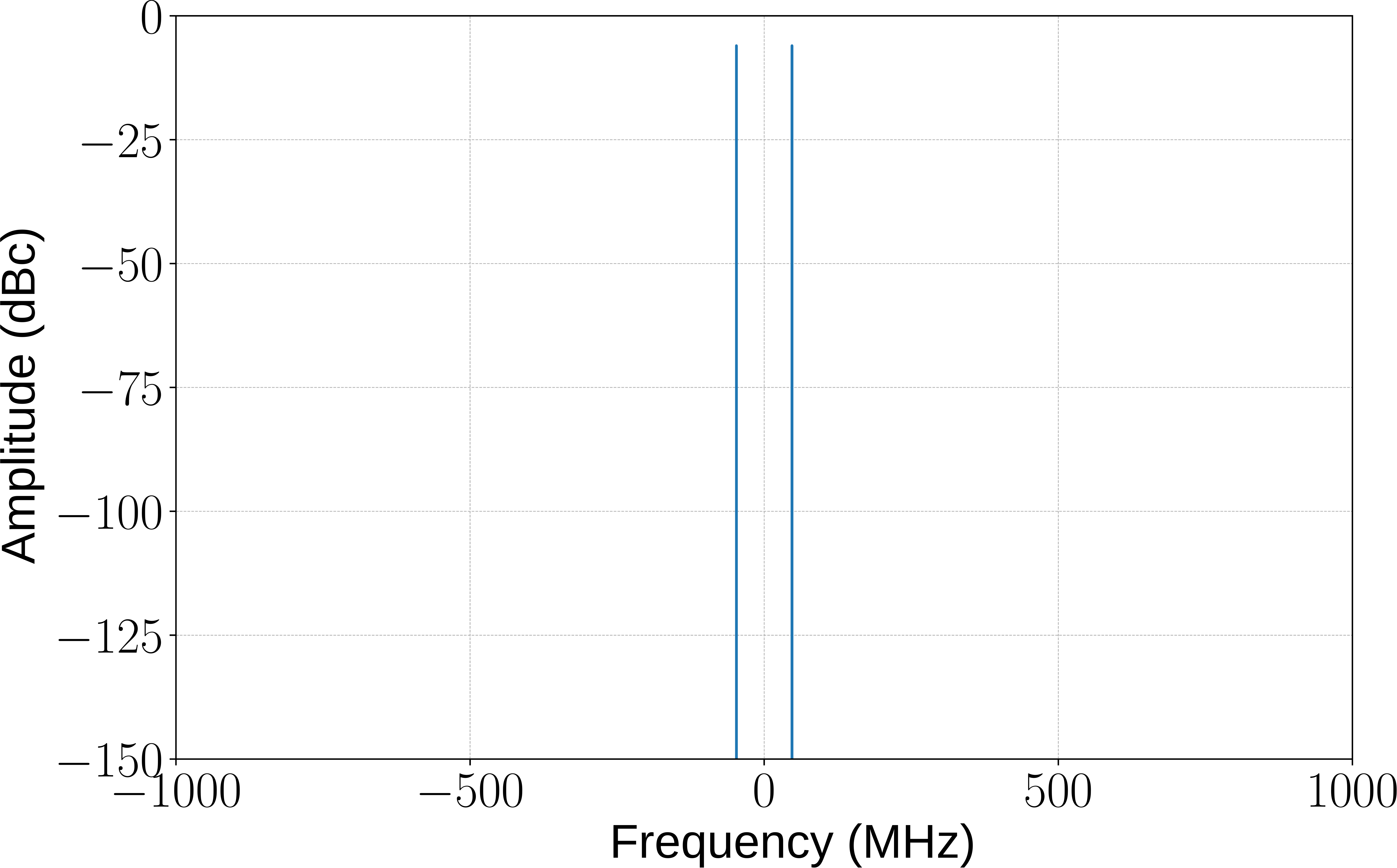}
        \caption{FFT with \(8 \times 2^{16}\) samples: sharp bin alignment.}
        \label{fig:pheno2a}
    \end{subfigure}
    \hfill
    \begin{subfigure}[t]{0.49\textwidth}
        \centering
        \includegraphics[width=\textwidth]{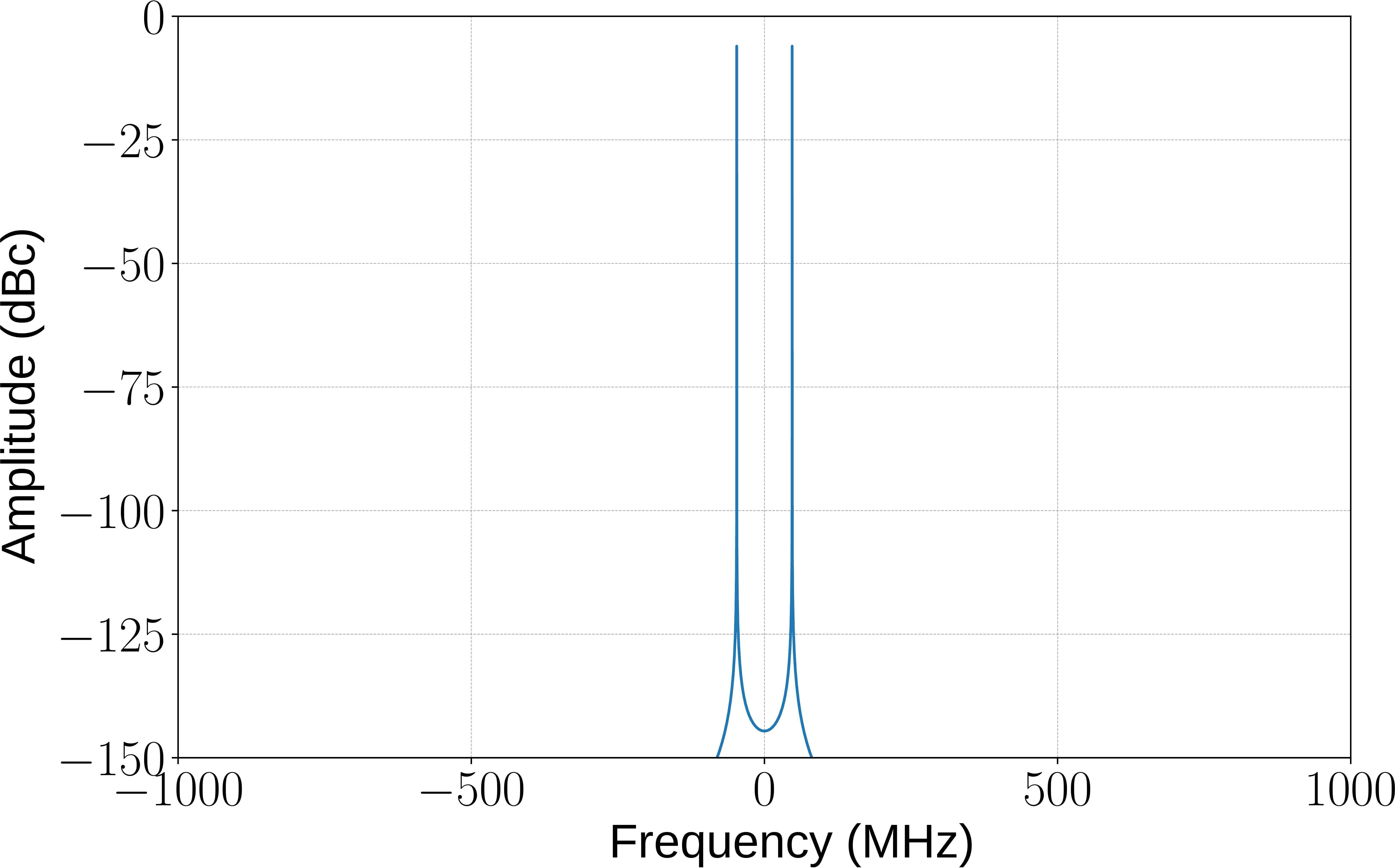}
        \caption{FFT with \(8 \times 2^{16} - 1\) samples: spectral broadening.}
        \label{fig:pheno2b}
    \end{subfigure}
    \caption{FFT spectra of the up-sampler output with sampling rate \(F_s = 2000\,\text{MHz}\).}
    \label{fig:pheno2}
\end{figure}

As illustrated in Fig.~\ref{fig:pheno2}, FFT over $8 \times 2^{16}$ samples shows distinct bin alignment, confirming the extended periodicity; spectral dispersion reappears when the FFT is computed over $8 \times 2^{16} - 1$ samples.

\subsubsection{Band-shifter}

After up-sampling, the I/Q signals are frequency-up-shifted using a complex exponential modulator that is specific to each band.  
This operation moves the signal spectrum from its centered range of \([-50\,\text{MHz},\,50\,\text{MHz}]\) into its designated output subband—for example, \([0\,\text{MHz},\,100\,\text{MHz}]\), \([100\,\text{MHz},\,200\,\text{MHz}]\), and so on, up to \([900\,\text{MHz},\,1000\,\text{MHz}]\).

This modulation is implemented in the up-shifter block  which uses a lookup table of 40~precomputed complex phasors (sine and cosine pairs).

For each band index \(i_{\text{band}} \in \{0, 1, \ldots, 9\}\), the shift is applied by a complex multiplication of the upsampled I/Q signals, \(I_{\text{up}}(n)\) and \(Q_{\text{up}}(n)\), by the complex phasor of the form:
\[
e^{j \alpha_{\text{iband}} n}, \quad \text{where} \quad \alpha_{\text{iband}} = \frac{2i_{\text{band}} + 1}{40} \cdot 2\pi.
\]

This corresponds to a frequency shift of:
\[
f_{\text{shift}} = \frac{\alpha_{\text{iband}}}{2\pi} \cdot F_s = \frac{2i_{\text{band}} + 1}{40} \cdot F_s.
\]

In this study, the band is arbitrarily chosen to be 6 as illustrated in Fig.~\ref{fig:blocks_anal}, which corresponds to a frequency shift of:
\[
f_{\text{shift}} = \frac{13}{40} \cdot 2000 = 650\,\text{MHz},
\]
which shifts the band from \([-50\,\text{MHz},\,50\,\text{MHz}]\) to \([600\,\text{MHz},\,700\,\text{MHz}]\).

To describe the modulation operation mathematically, we first express the upsampled in-phase and quadrature signals as a single complex-valued signal:
\[
x_{\text{up}}(n) = I_{\text{up}}(n) + j Q_{\text{up}}(n).
\]
The modulation is then applied by multiplying this signal by the complex exponential:
\[
x_{\text{shifted}}(n) = x_{\text{up}}(n) \cdot e^{j \alpha_{\text{iband}} n} = \left[ I_{\text{up}}(n) + j Q_{\text{up}}(n) \right] \cdot \left[ \cos(\alpha_{\text{iband}} n) + j \sin(\alpha_{\text{iband}} n) \right].
\]

Separating the result into real and imaginary components yields the shifted real I/Q outputs:
\[
\begin{aligned}
I_{\text{shifted}}(n) &= I_{\text{up}}(n) \cos\left( \alpha_{\text{iband}} n \right) - Q_{\text{up}}(n) \sin\left( \alpha_{\text{iband}} n \right), \\
Q_{\text{shifted}}(n) &= Q_{\text{up}}(n) \cos\left( \alpha_{\text{iband}} n \right) + I_{\text{up}}(n) \sin\left( \alpha_{\text{iband}} n \right).
\end{aligned}
\]

\paragraph{Time Domain:} 

From a sample-domain perspective, the introduction of the modulator modifies the periodicity.
Specifically, the signals \( I_{\text{shifted}}(n) \) and \( Q_{\text{shifted}}(n) \) no longer repeat every \(8 \times 2^{16}\) samples.  
Instead, the new periodicity is determined by LCM of the upsampled tone periodicity \((8 \times 2^{16})\) and the modulator’s periodicity (40 samples).  
Since

\[
\text{LCM}(8 \times 2^{16}, 40) = 5 \times 8 \times 2^{16}= 5 \times 2^{19},
\]

the overall periodicity becomes \(5 \times 2^{19}\) samples, reflecting the combined effects of upsampling and modulation.
We compared multiple consecutive \(5 \times 2^{19}\) segments, and confirmed the combined, new, periodicity :

\[
I_{\text{shifted}}[n] = I_{\text{shifted}}[n + 5 \times 2^{19}], \quad 
Q_{\text{shifted}}[n] = Q_{\text{shifted}}[n + 5 \times 2^{19}] \quad \forall n
\]

\paragraph{Frequency domain:} 

\begin{figure}[h]
    \centering
    \begin{subfigure}[t]{0.49\textwidth}
        \centering
        \includegraphics[width=\textwidth]{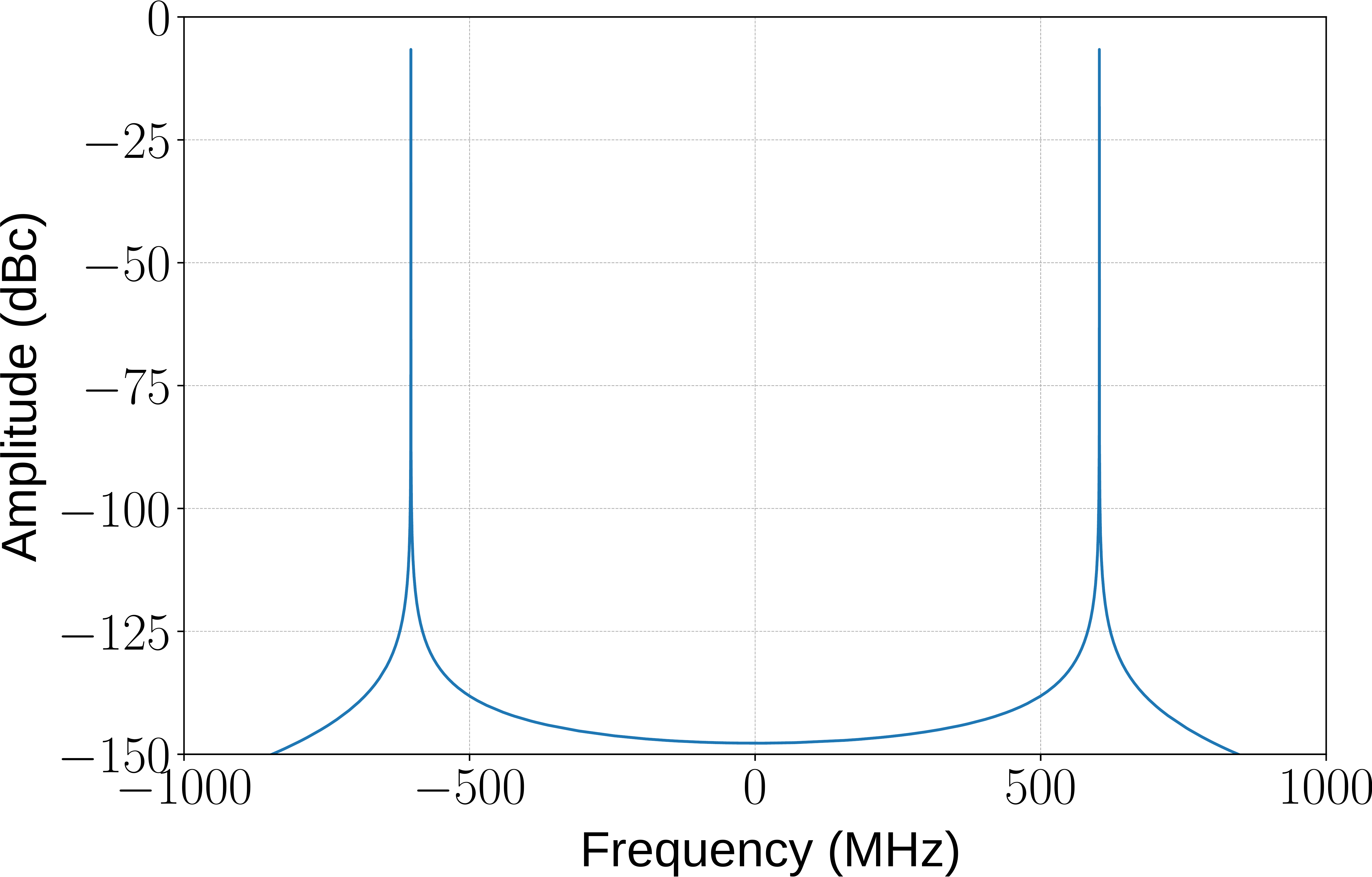}
        \caption{FFT with \(8 \times 2^{16}\) samples: spectral broadening.}
        \label{fig:pheno3a}
    \end{subfigure}
    \hfill
    \begin{subfigure}[t]{0.49\textwidth}
        \centering
        \includegraphics[width=\textwidth]{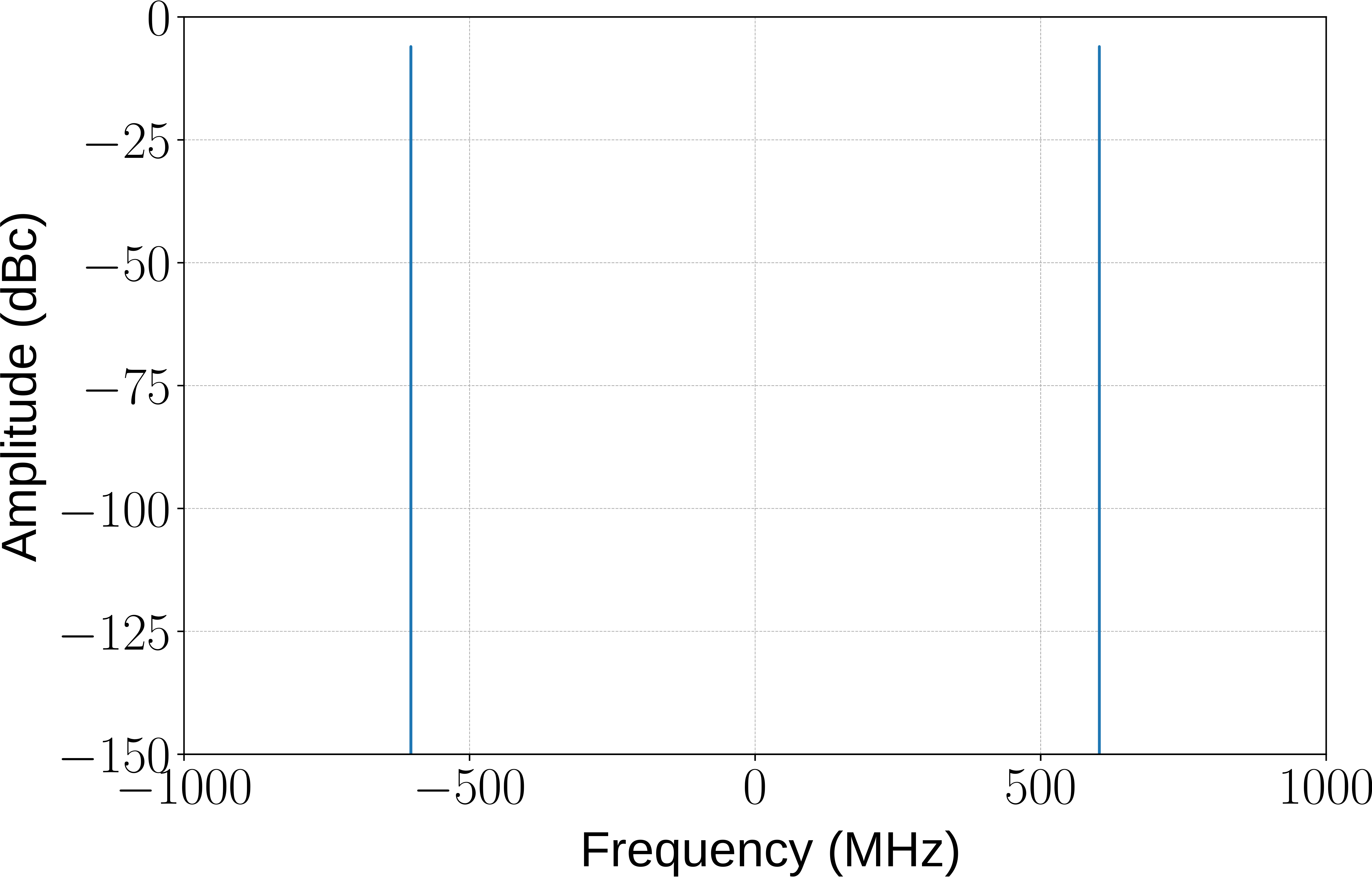}
        \caption{FFT with \(5 \times 8 \times 2^{16}\) samples: sharp bin alignment.}
        \label{fig:pheno3b}
    \end{subfigure}
    \caption{FFT spectra of the up-shifter output with a sampling rate of \(F_s = 2000\,\text{MHz}\).  
The baseband tone at \(f_1 = -47.24\,\text{MHz}\) is up-shifted by \(f_2 = 650\,\text{MHz}\) to band 6, resulting in a tone located at \(f = 602.76\,\text{MHz}\).}
    \label{fig:pheno3}
\end{figure}

Fig.~\ref{fig:pheno3} demonstrates that the previous FFT size of \(8 \times 2^{16}\) causes spectral dispersion.
However, the \(5 \times 8 \times 2^{16}\)-point FFT restores bin alignment, confirming the new \(5 \times 2^{19}\) periodicity.

\subsubsection{Polyphase Filter}
\label{poly_imper}

The polyphase filter implemented in the FPGA is fed with a real-valued signal—specifically, the return signal from the MKID array, digitized by the ADC. However, in this investigation, we loop back the excitation signal. Since the excitation signal is in quadrature, the loopback is performed by feeding only one component (either the In-phase or Quadrature signal) from the band-shifter back into the polyphase filter, as illustrated in Fig.~\ref{fig:blocks_anal}.

The polyphase filter performs five key operations: (1) complex down-shifting, (2) filtering to isolate the sub-bands, (3) down-sampling from 2000\,MHz to 250\,MHz, (4) up-conversion by +62.5\,MHz , and (5) conversion of the resulting complex signal back to a real-valued one.
In this study, we focus primarily on the demodulation step, which is directly related to the spectral artifacts under investigation.

The demodulation is carried out using a complex exponential with frequency opposite to that applied during the up-shifting stage of the excitation path by up-shifter.
For each band \( i_{\text{band}} \), the demodulation is performed using a phasor:

\[
e^{-j \alpha_{\text{iband}} n}, \quad \text{where} \quad \alpha_{\text{iband}} = \frac{2i_{\text{band}} + 1}{40} \cdot 2\pi.
\]

We recall that the input to the polyphase filter is a real-valued signal.
Consequently, the demodulation performed at this stage consists of multiplying a real-valued signal by a complex phasor.
This differs from the excitation chain, where both the signal and the modulation are complex-valued.

The purpose of this demodulation step is to center only the positive frequency content of the target subband around \([-50\,\text{MHz}, +50\,\text{MHz}]\).

\paragraph{Frequency Domain:} 

\begin{figure}[h]
    \centering
    \begin{subfigure}[t]{0.49\textwidth}
        \centering
        \includegraphics[width=\textwidth]{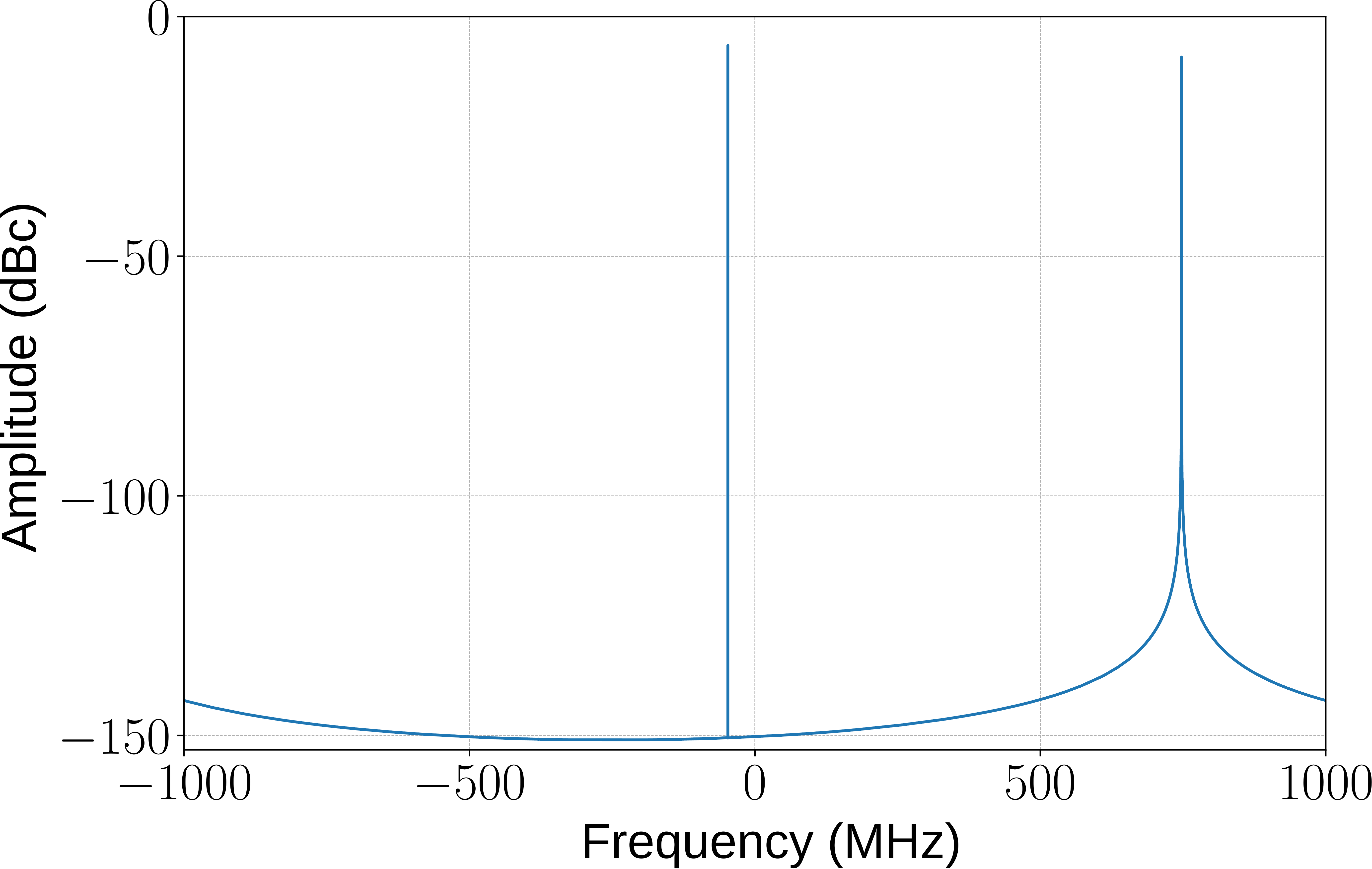}
        \caption{FFT with \(8 \times 2^{16}\) samples: spectral broadening.}
        \label{fig:pheno4a}
    \end{subfigure}
    \hfill
    \begin{subfigure}[t]{0.49\textwidth}
        \centering
        \includegraphics[width=\textwidth]{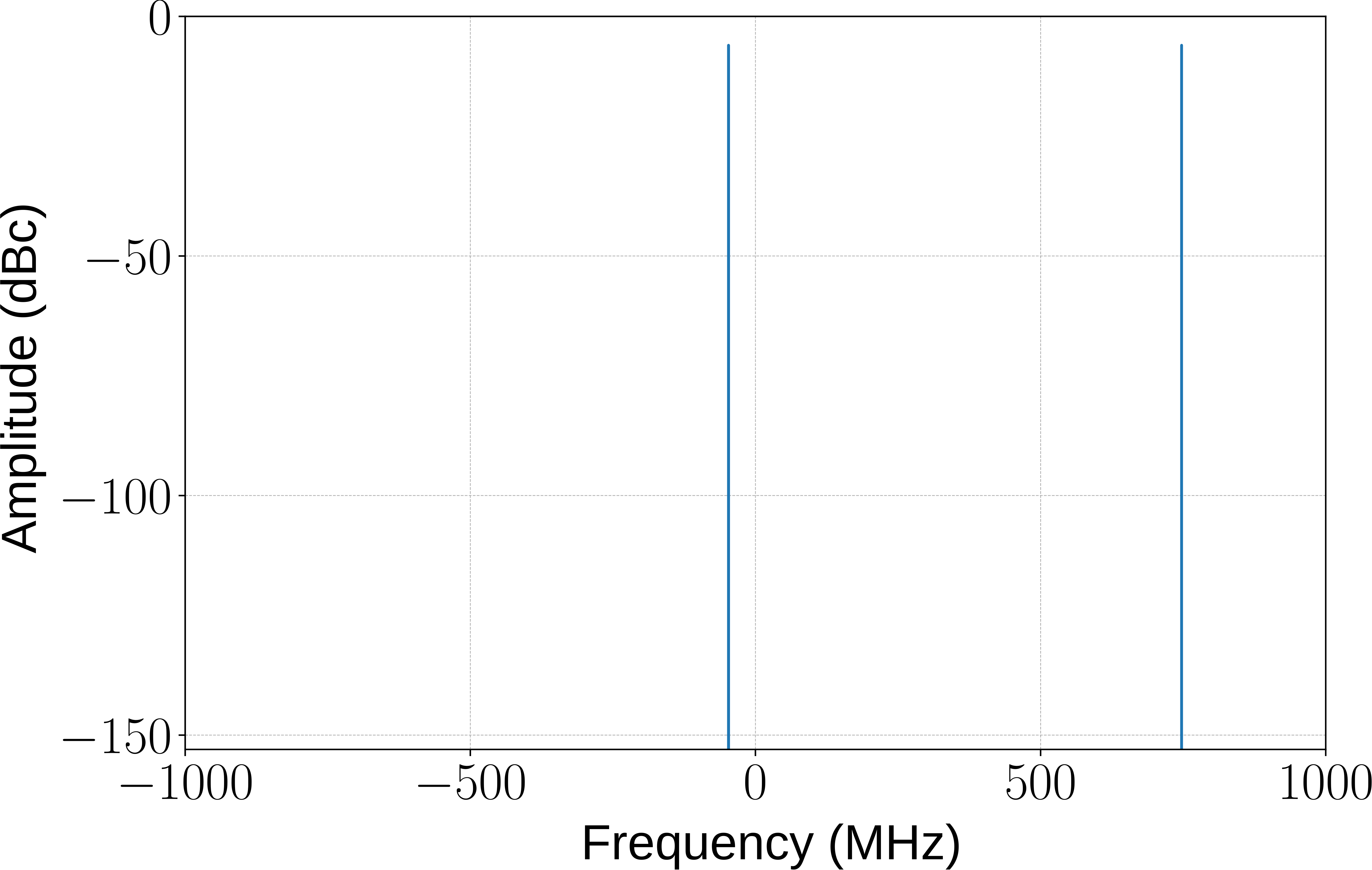}
        \caption{FFT with \(5 \times 8 \times 2^{16}\) samples: sharp bin alignment.}
        \label{fig:pheno4b}
    \end{subfigure}
    \caption{FFT spectra of the polyphase demodulator output with sampling rate \(F_s = 2000\,\text{MHz}\).}
    \label{fig:pheno4}
\end{figure}

The demodulation phasor at frequency \(-f_2 = -650\,\text{MHz}\) shifts all spectral components downward by 650\,MHz.  
As shown in Fig.~\ref{fig:pheno4}, the original tone—previously up-shifted to \(f = 602.76\,\text{MHz}\) (i.e., \(f_1 - 62.5 + f_2\))—is now shifted back to:
\[
f = 602.76 - 650 = -47.24\,\text{MHz},
\]
which falls within the baseband range of \([-50\,\text{MHz}, +50\,\text{MHz}]\).

As for  corresponding negative-frequency component is located at:
\[
f = -(f_1 - 62.5 + f_2) = -602.76\,\text{MHz}.
\]
It is translated to:
\[
f = -(f_1 - 62.5 + f_2) - f_2 = -(f_1 - 62.5 + 2f_2) = -1252.76\,\text{MHz}.
\]

This lies outside the Nyquist range of \(\pm1000\,\text{MHz}\) (given \(F_s = 2000\,\text{MHz}\)), which folds back into the observable spectrum as:
\[
f_{\text{alias}} = 2000 - 1252.76 = 747.24\,\text{MHz},
\]
as illustrated in Fig.~\ref{fig:pheno4}.

However, $f_2$ is still present in the \(-(f_1 - 62.5 + 2f_2)\) component, and since it originates from the up-shifter, the resulting demodulated signal still retains the underlying periodicity of \(5 \times 2^{19}\) samples.
This behavior is illustrated in Fig.~\ref{fig:pheno4}\,(a): when a Fourier transform of size $2^{16} \times 8$ is applied, the positive-frequency component at $f_1$ aligns perfectly with a discrete Fourier bin. 
In contrast, the negative-frequency component \(-(f_1 - 62.5 + 2f_2)\) appears spectrally dispersed, providing clear evidence of the preserved \(5 \times 2^{19}\)-sample periodicity, as demonstrated by the perfect spectral alignment observed with the $2^{16} \times 8 \times 5$ sample length FFT in Fig.~\ref{fig:pheno4}\,(b).

While the polyphase filter provides approximately \(-80\,\text{dB}\) of stopband attenuation beyond its 50\,MHz low-pass cutoff, this level of rejection has proven insufficient, as spurious tones are observed at 763\,MHz and 1526\,MHz at the DDC output.  
This issue will be discussed in the next section.

The final steps of the polyphase filtering process have no impact on the signal’s periodicity.  
The +62.5\,MHz up-conversion is performed using the same phasor applied during excitation but with the opposite sign, \( e^{j\pi n / 2} \).  
Since this phasor has a periodicity of 4~samples, and \(\text{LCM}(5 \times 2^{19}, 4) = 5 \times 2^{19}\), the overall periodicity of \(5 \times 2^{19}\) samples is preserved.  

\subsubsection{Tone analyzer:}

The tone analyzer consists of a DDC that performs three operations.
It first demodulates the sub-band signal, produced by the polyphase filter, by multiplying it with the original quadrature excitation tone $f_1$, as illustrated in Fig.~\ref{fig:blocks_anal}. The resulting signal then passes through an LPF before being downsampled from 250\,MHz to 3.8\,kHz.

The LPF used in the DDC is an averaging filter with a window length of \(2^{16}\), deliberately chosen to match the length of the phase accumulator in the tone generator. 
This design choice has a significant consequence: the averaging filter's frequency response exhibits zeros at integer multiples of \(\Delta f = 250\,\mathrm{MHz} / 2^{16}\). 
Interestingly, this zero spacing is identical to the frequency resolution of the tone generator.
As a result, after demodulation of $f_1$ to 0\,Hz, all other demodulated tones are suppressed.

\paragraph{Frequency Domain}

The 763\,Hz and 1526\,Hz observed in the output spectra of all DDCs (see Fig.~\ref{fig:simu_resu_digital}) originate from negative-frequency components introduced during complex demodulation within the polyphase filter structure—specifically, terms of the form \(-(f_1 - 62.5 + 2f_2)\) discussed in Section~\ref{poly_imper}.

In the FFT spectrum shown previously (Fig.~\ref{fig:pheno4}), these components appear dispersed because their frequencies do not align with the discrete FFT bins, which are spaced by \(\frac{8 \times 250\,\text{MHz}}{8 \times 2^{16}}\). 
This bin spacing also corresponds to the locations of the zeros of the averaging filter, i.e., \(\frac{250\,\text{MHz}}{2^{16}}\).
As a result, the spurious components do not fall on these nulls and are therefore not completely rejected.

These misaligned frequency components subsequently leak through the filter and alias into the baseband during downsampling, ultimately manifesting as the observed spurious tones at 763\,Hz and 1526\,Hz.

\subsection{Proposed solution}
\label{sec:proposed_solution_65500}
To mitigate the spurious components, we propose two complementary modifications targeting both the digital excitation and analysis chains.

First, within the excitation chain, the current 16-bit phase accumulator—operating modulo $2^{16}$—is replaced with one having a period of 65,520. 
This change is designed to alter the, previously discussed, LCM relationships that govern signal periodicity across subsequent processing stages.

Second, in the analysis chain, the averaging filter within the DDC is modified. 
Instead of averaging over $2^{16}$ samples, it now operates over 65520 samples. 
This adjustment ensures that the filter notches align with the spectral positions of spurious components resulting from extended periodicities introduced earlier in the chain.

\subsubsection{Impact on Excitation Chain}

By adopting a phase accumulation period of 65520~samples, three main effects arise.

First, the frequency resolution changes from:

\[
\frac{250\,\text{MHz}}{2^{16}} \approx 3814.697\,\text{Hz} \quad \text{to} \quad \frac{250\,\text{MHz}}{65520} \approx 3815.628\,\text{Hz}.
\]

This represents a frequency resolution degradation of approximately 1\,Hz, which is negligible and does not impact the ability to accurately probe the resonators.

Second, although the phase accumulator's output range now spans 0 to 65519 instead of 0 to $2^{16}-1$, this change has no impact on the CORDIC. 
This because in the implemented design, the accumulator's output undergoes a 6-bit right shift before reaching the CORDIC, which receives only the 10 most significant bits (MSBs). Consequently, the missing values in the range 65520–65535---which differ only in their least significant 4 bits that are discarded anyway---produce identical 10-bit MSBs fed to the CORDIC input. 

Third, and most critically, the choice of 65520 is intentional: it is divisible by 40, the periodicity of phasor used for frequency shifting in the up-shifting stage. 
As a result, the overall signal periodicity—considering upsampling by a factor of 8—remains unchanged:

\[
\text{LCM}(8 \times 65520,\ 40) = 8 \times 65520,
\]

which eliminates the fivefold increase in periodicity observed in the previous configuration with a 16-bit accumulator. Digital twin simulations confirm this at the up-shifter output:

\[
I_{\text{shifted}}[n] = I_{\text{shifted}}[n + 8 \times 65520], \quad 
Q_{\text{shifted}}[n] = Q_{\text{shifted}}[n + 8 \times 65520], \quad \forall n.
\]

In the frequency domain, $8 \times 65520$-point FFTs yield perfect bin alignment---unlike the spectral broadening with prior $8 \times 2^{16}$-point FFTs (Fig.~\ref{fig:pheno3}).

\subsubsection{Impact on Polyphase Demodulation}

This modification's benefit is evident at the polyphase demodulator output. The problematic negative-frequency term $-(f_1 - 62.5 + 2f_2)$, previously misaligned and spectrally dispersed (Fig.~\ref{fig:pheno4}), now aligns perfectly with $8 \times 65520$-point FFT bins spaced at $\frac{8 \times 250\,\mathrm{MHz}}{8 \times 65520} = \frac{250\,\mathrm{MHz}}{65520}$. These coincide exactly with the averaging filter's nulls, suppressing the spurious components effectively.

\subsubsection{Impact on DDC Output}

\begin{figure}[H]
    \centering

    \begin{subfigure}[t]{0.98\textwidth}
        \centering
        \begin{subfigure}[t]{0.48\textwidth}
            \centering
            \includegraphics[width=\textwidth]{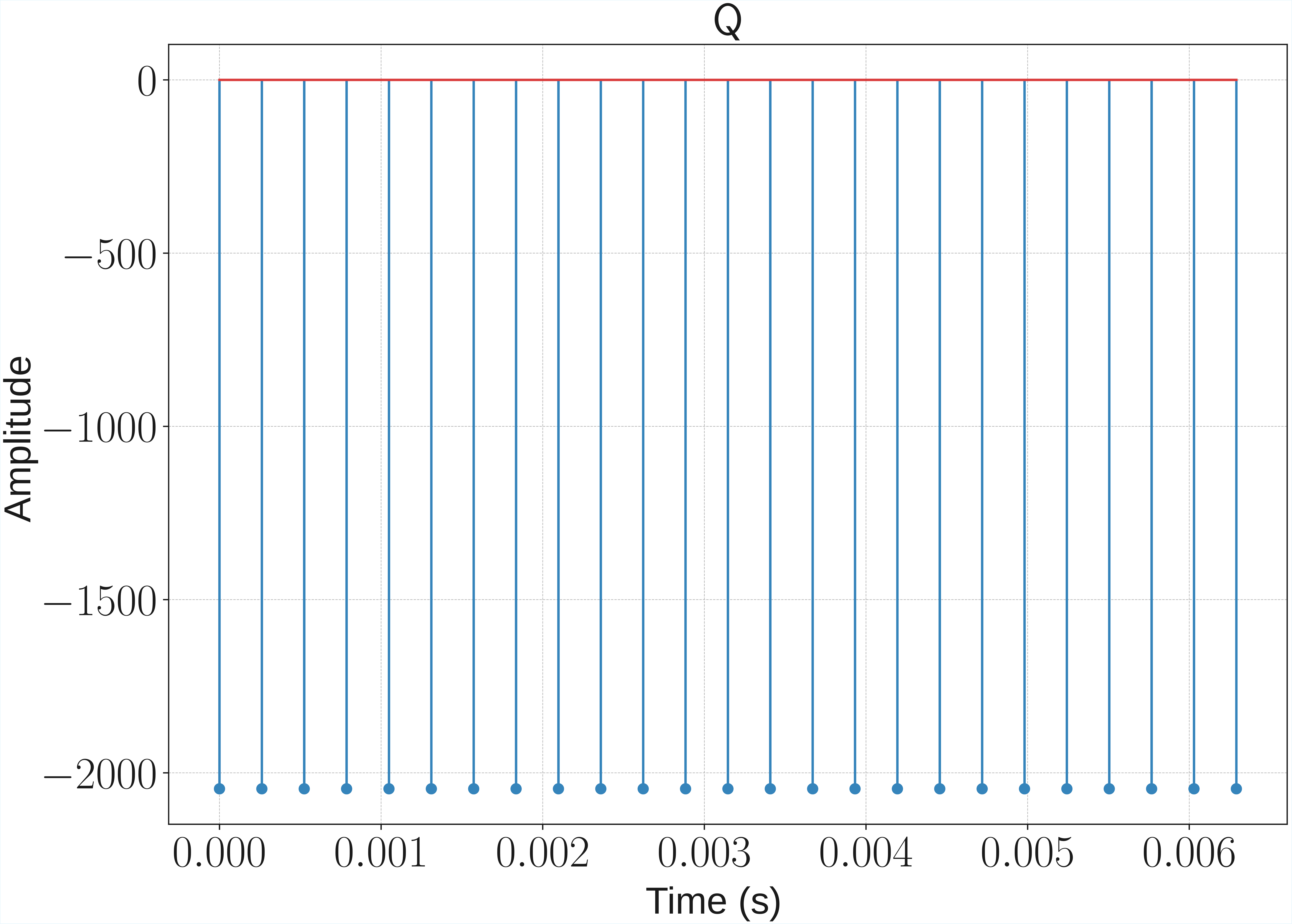}
        \end{subfigure}
        \hfill
        \begin{subfigure}[t]{0.48\textwidth}
            \centering
            \includegraphics[width=\textwidth]{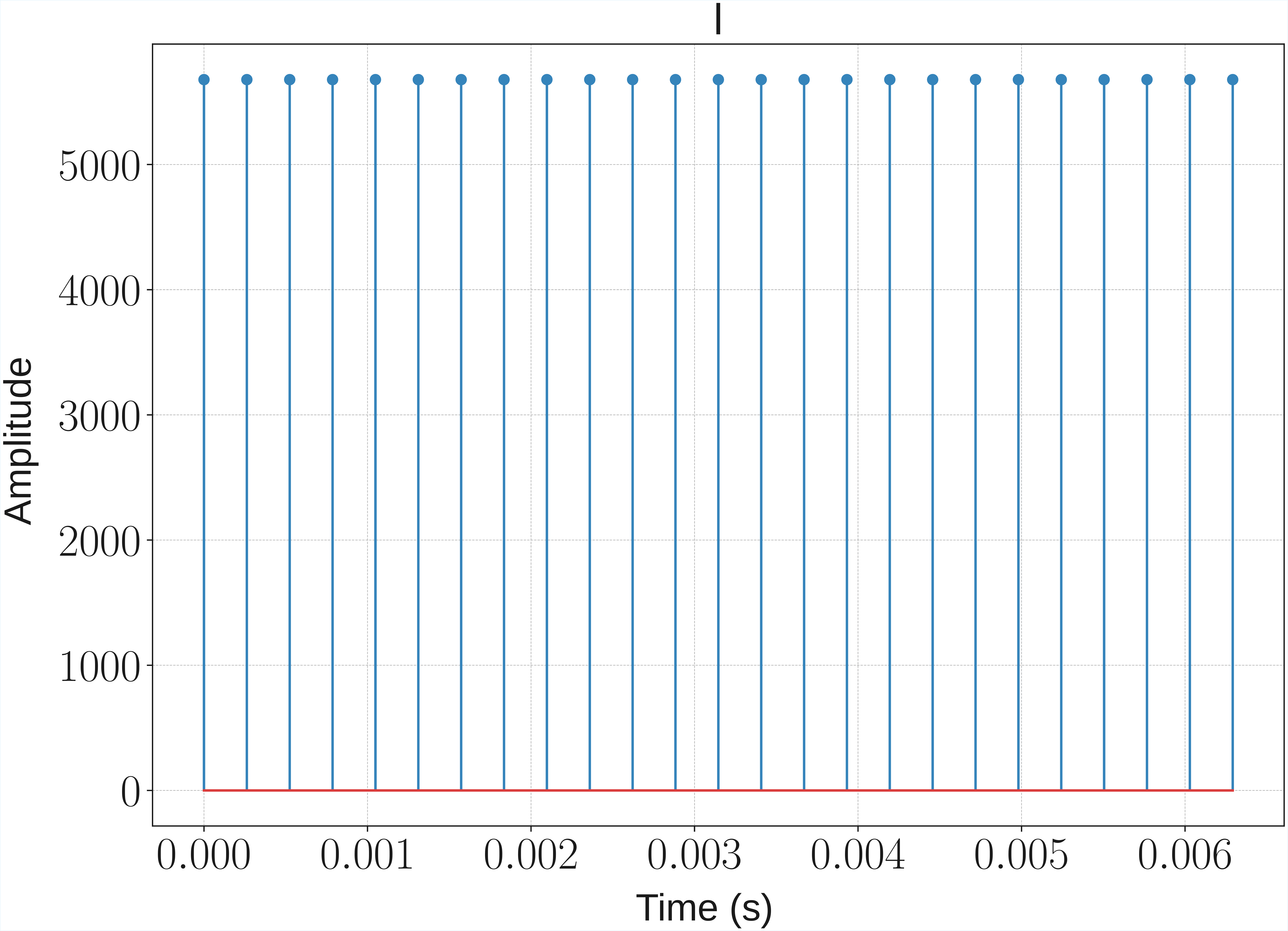}
        \end{subfigure}
        \caption{}
        \label{fig:setup1a}
    \end{subfigure}

    \vspace{0.3cm}

    \begin{subfigure}[t]{0.98\textwidth}
        \centering
        \begin{subfigure}[t]{0.48\textwidth}
            \centering
            \includegraphics[width=\textwidth]{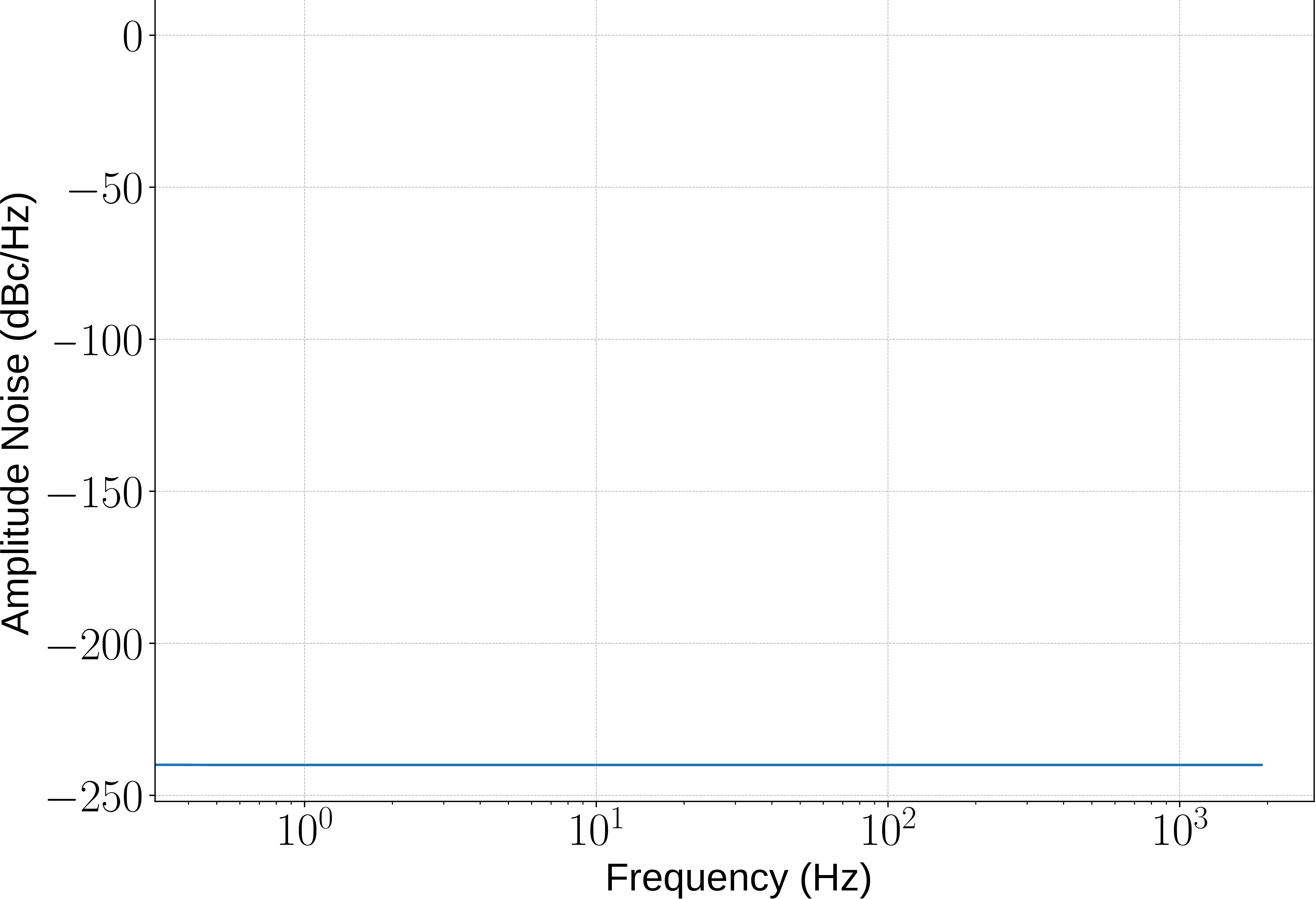}
        \end{subfigure}
        \hfill
        \begin{subfigure}[t]{0.48\textwidth}
            \centering
            \includegraphics[width=\textwidth]{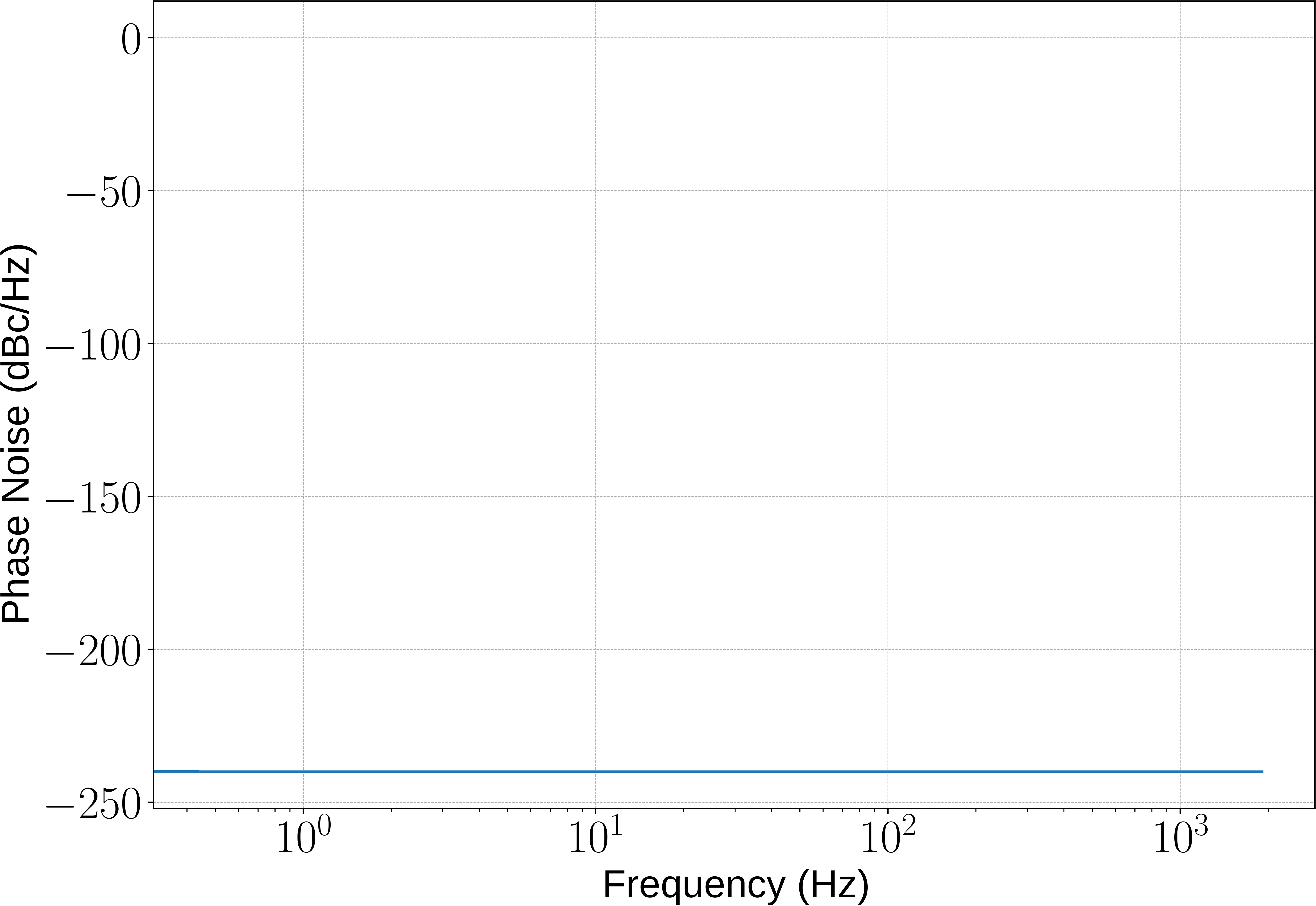}
        \end{subfigure}
        \caption{}
        \label{fig:setup1b}
    \end{subfigure}

\caption{Simulation results: (a) Time-domain I/Q signals; (b) corresponding amplitude and phase noise PSD.}

    \label{fig:proposed}
\end{figure}

The effectiveness of the proposed strategy is confirmed by examining the I/Q signals at the DDC output. 
As illustrated in Fig.~\ref{fig:proposed}\,(a), in time domain, these signals are constant, indicating that the DDC has demodulated the tone to DC, while all other spectral components have been effectively rejected by the averaging filter.
In the frequency domain, the amplitude and phase noise PSD show no spurious tones—demonstrating that the proposed modifications effectively eliminate the 763\,MHz and 1526\,MHz, as illustrated in Fig.~\ref{fig:proposed}\,(b).

\subsubsection{Output sampling rate Impact}

Finally, it is important to note that the downsampling performed at the output of each DDC is now carried out by a factor of 65520 instead of \(2^{16}\).  
As a result output sampling rate is reduced from \(250\,\text{MHz}\) to approximately \(3.81562\,\text{kHz}\), replacing the previous rate of \(3.81469\,\text{kHz}\). 
The difference between the two rates is negligible and does not affect the temporal resolution required for the CONCERTO interferometric experiments.


\subsection{Implementation and Measurement}

\subsubsection{VHDL modifications}

Initially, the phase accumulator was implemented as a 16-bit incrementer: the phase was updated as \texttt{phase} $\leftarrow$ \texttt{phase} $+$ \texttt{FCW}, and naturally wrapped around due to the register's fixed bit width. This behavior implicitly performed a modulo-$2^{16}$ operation.

To implement the new accumulation period, this logic was replaced by an explicit modulo operation:
\[
\texttt{phase} \leftarrow (\texttt{phase} + \texttt{FCW}) \bmod 65520.
\]

\subsubsection{Measurement Setup}

\begin{figure}[H]
    \centering
    \includegraphics[width=1\textwidth]{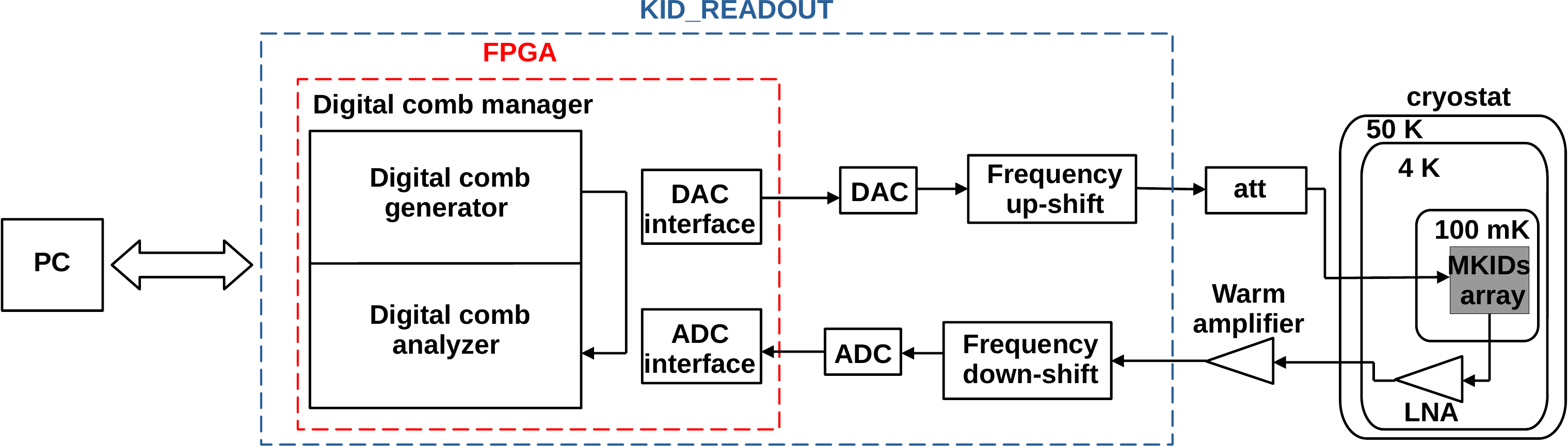}
    \caption{Block diagram of the setup.}
    \label{fig:setups}
\end{figure}

The VHDL firmware was modified to create a purely digital loop-back from the digital comb generator to the digital comb analyzer, as illustrated in Fig.~\ref{fig:setups}, bypassing the remainder of the readout chain.
This modification was motivated by the digital twin, which demonstrated that the source of the spurious tones is entirely digital.

Using this digital loop-back configuration, two firmware versions were implemented.
The first uses the original $2^{16}$ phase accumulator together with the DDC window-averaging filter, while the second adopts the proposed $65{,}520$-based implementation.

For the measurements, 400 I/Q data streams were acquired at system sampling rates of approximately $3.81562\,\text{kHz}$ for the modified firmware and $3.81468\,\text{kHz}$ for the original firmware, over a duration of about 3~minutes.
In both cases, the acquisition time was selected to yield the same number of samples per tone, namely $N = 655{,}360$, ensuring consistent and sufficiently fine frequency resolution for detailed spectral analysis.

\subsubsection{Measurement analysis}  

\begin{figure}[H]
    \centering

    \begin{subfigure}[t]{0.98\textwidth}
        \centering
        \begin{subfigure}[t]{0.48\textwidth}
            \centering
            \includegraphics[width=\textwidth]{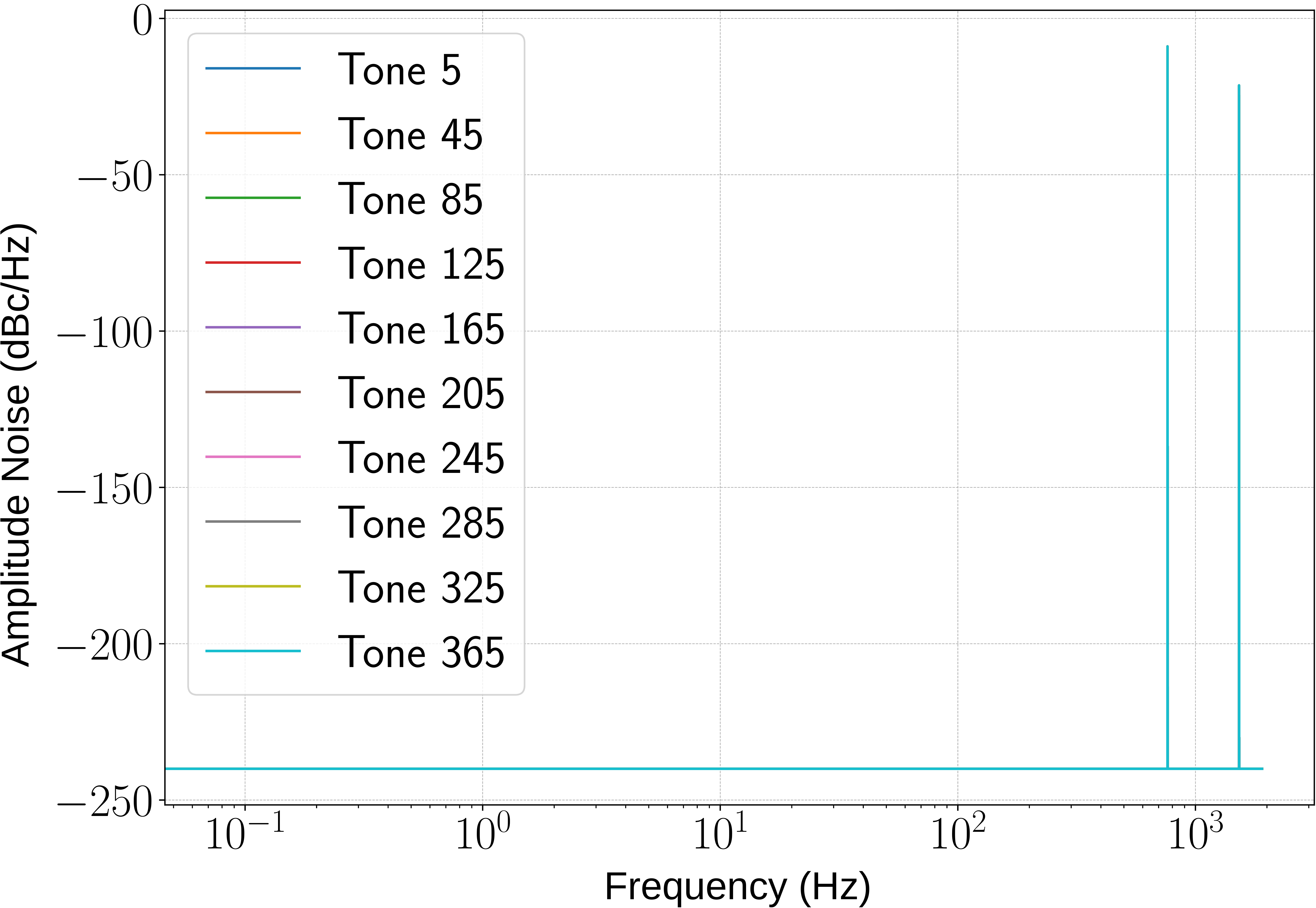}
        \end{subfigure}
        \hfill
        \begin{subfigure}[t]{0.48\textwidth}
            \centering
            \includegraphics[width=\textwidth]{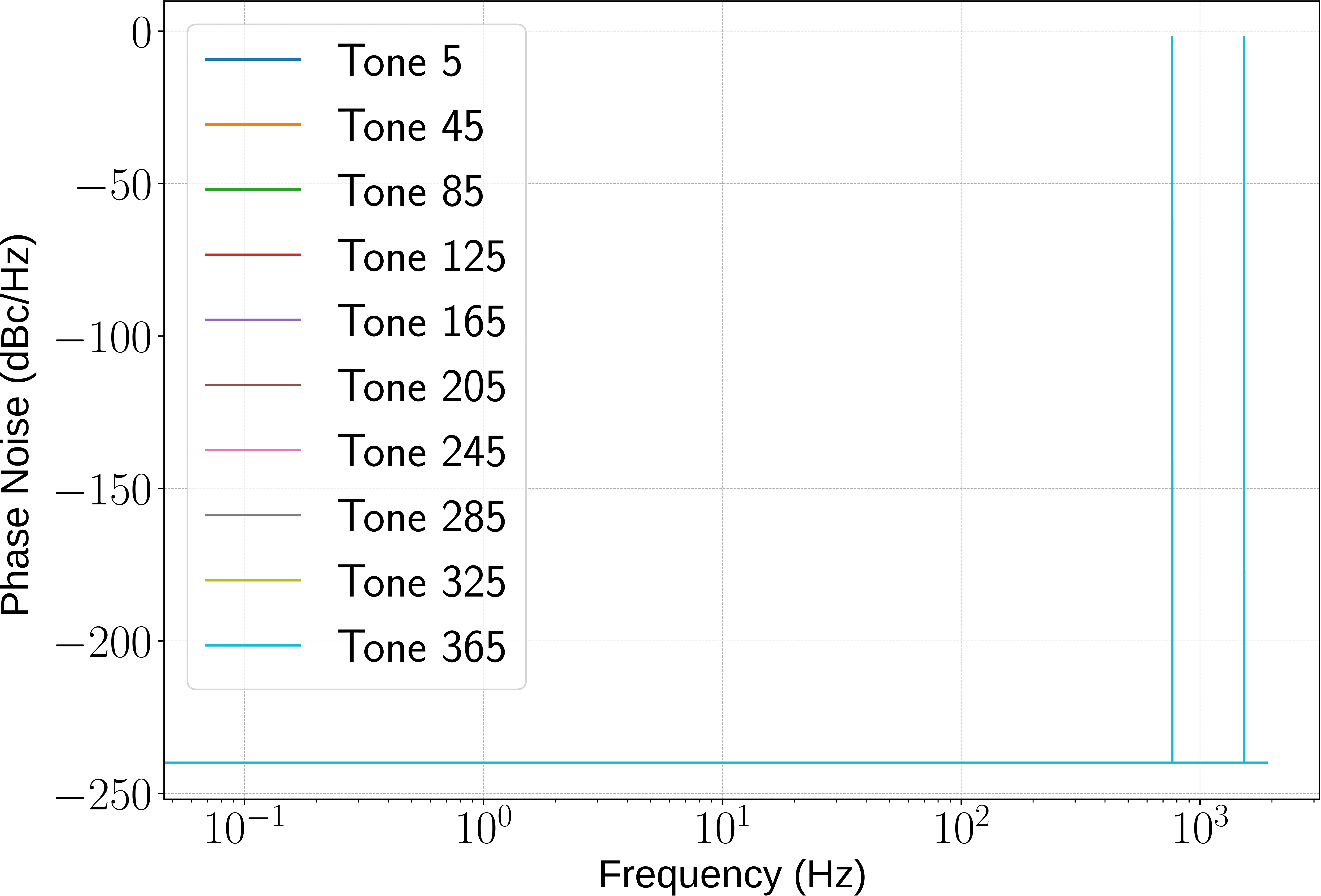}
        \end{subfigure}
        \caption{}
        \label{fig:setup1a}
    \end{subfigure}

    \vspace{0.3cm}

    \begin{subfigure}[t]{0.98\textwidth}
        \centering
        \begin{subfigure}[t]{0.48\textwidth}
            \centering
            \includegraphics[width=\textwidth]{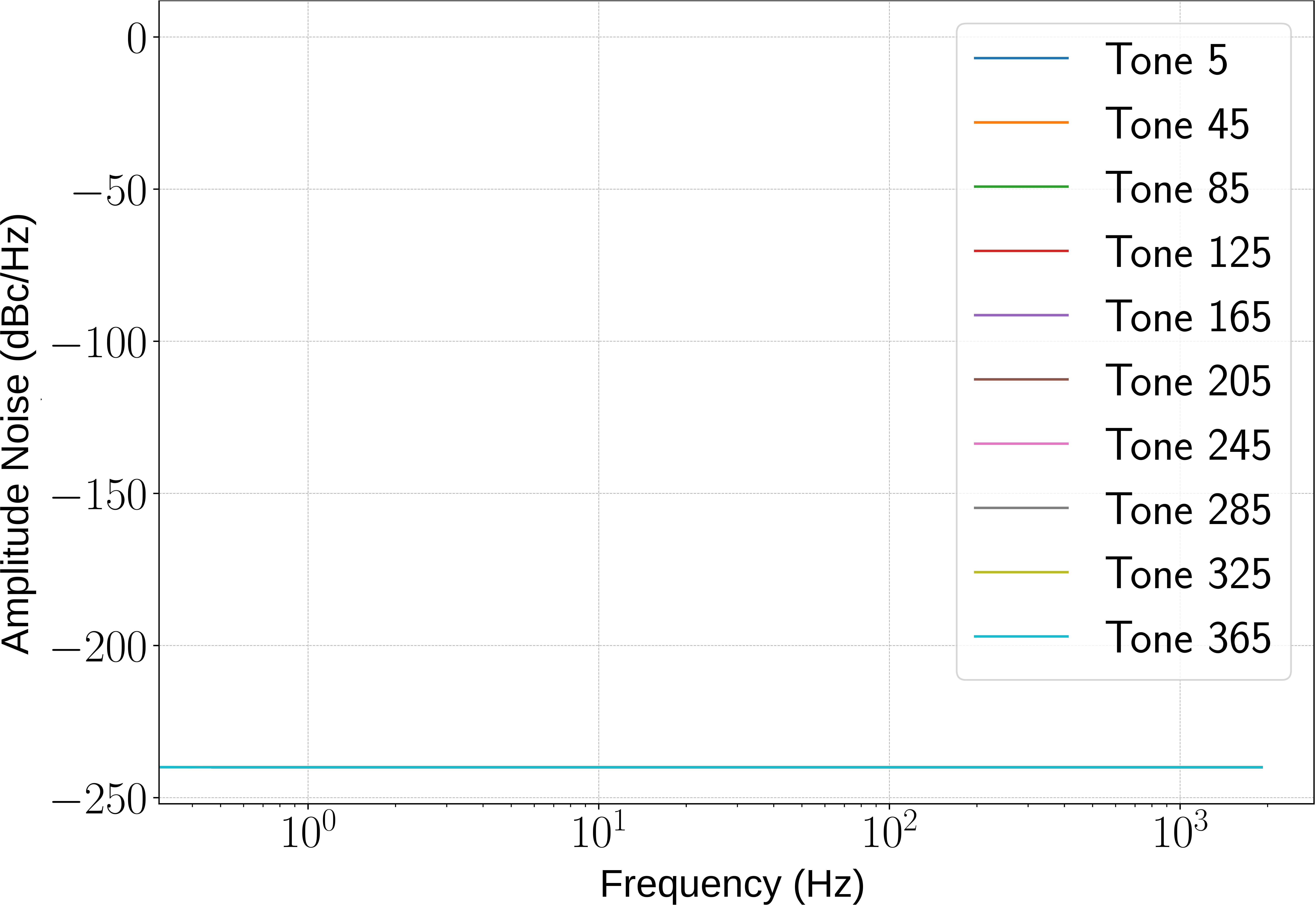}
        \end{subfigure}
        \hfill
        \begin{subfigure}[t]{0.48\textwidth}
            \centering
            \includegraphics[width=\textwidth]{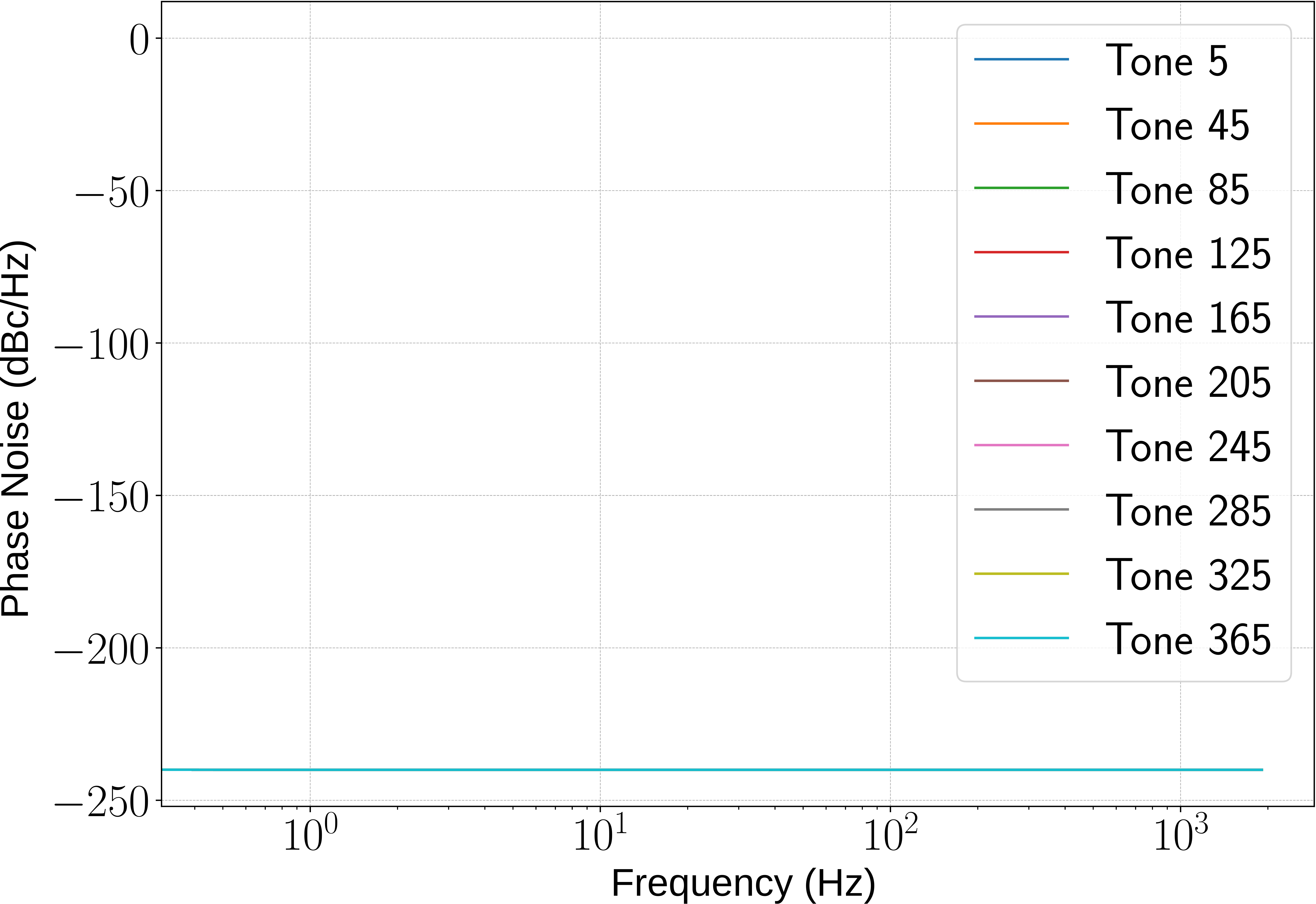}
        \end{subfigure}
        \caption{}
        \label{fig:setup1b}
    \end{subfigure}

\caption{Comparison of measured amplitude and phase noise PSD, between the original firmware (a) and the proposed modified firmware (b). 
A representative tone from each of the 10 frequency bands is shown.}

    \label{fig:setup1_all}
\end{figure}

With the original firmware, the measured spectra exhibit two distinct peaks at 763\,Hz and 1526\,Hz (see Fig.~\ref{fig:setup1_all}(a)), reaffirming that the phenomenon originates within the digital processing chain.

In contrast, Fig.~\ref{fig:setup1_all}(b) shows that, when using the modified firmware, both the amplitude and phase noise spectra display a flat noise floor at $-240\,\text{dBc/Hz}$. 
This confirms that the DDC performs demodulation to DC followed by effective low-pass filtering. 
These results are in excellent agreement with the simulation outcomes presented in Section~\ref{sec:proposed_solution_65500}.

\subsubsection{FPGA resource utilization}

In the original implementation, the accumulator naturally wrapped at \(2^{16}\) due to the fixed register width, effectively performing a modulo-\(2^{16}\) operation.  
However, changing this to a modulus of 65520, which is a non-power-of-two constant, demands more resources.

Table~\ref{tab:resource_usage_65520} summarizes the FPGA resource utilization comparison between the complete readout firmware implementations on the Xilinx XCKU060FFVA1156-2 FPGA.

For reference, the total available FPGA resources are as follows: LUTs: 331 680, Flip-Flops (FFs): 663 360, DSPs: 2760.
 
The 65520-based version increases LUT usage by 10758 LUT (from 215228 to 225986),  a 3.24\%pt growth (from 64.89\% to 68.13\%), and FF usage increases by 26,520 FFs (from 346539 to 346759), a 0.03\%pt growth (from 52.24\% to 52.27\%)—while DSP usage remains unchanged.

\begin{table}[h]
\centering
\caption{FPGA resource utilization for the original and new firmware.}
\small
\begin{tabular}{|p{2.5cm}|p{1.5cm}|p{1.5cm}|p{1.5cm}|p{1.5cm}|p{1.2cm}|p{1.2cm}|}
    \hline
    \textbf{Accumulator/ Averaging Length} & 
    \textbf{CLB LUTs Count} & \textbf{CLB LUTs \% Used} & 
    \textbf{CLB FFs Count} & \textbf{CLB FFs \% Used} & 
    \textbf{DSPs Count} & \textbf{DSPs \% Used} \\
    \hline
    \(2^{16} = 65536\) & 215{,}228  & 64.89\% & 346{,}539 & 52.24\% & 1{,}953 & 70.76\% \\
    \hline
    65{,}520           & 225{,}986  & 68.13\% & 346{,}759 & 52.27\% & 1{,}953 & 70.76\% \\
    \hline
\end{tabular}
\label{tab:resource_usage_65520}
\end{table}

Despite a slight increase in LUT and FF utilization, the modified design remains well within the available resources of the target FPGA, making this trade-off acceptable given the significant spectral improvements achieved.

\section{Conclusion}

We presented, in this work, a cycle- and bit-accurate Python digital twin of the CONCERTO KID\_READOUT FPGA DSP chain and used it to investigate previously unexplained spurious components observed in the MKID readout.
The model reproduces the measured spurs at 763\,Hz and 1526\,Hz and enables stage-by-stage analysis, which identifies their origin as the combined effect of (i) a periodicity extension introduced by the band up-shifter and (ii) incomplete suppression of negative-frequency terms in the analysis path, which subsequently alias into baseband.
Guided by these findings, we implemented a mitigation strategy based on replacing the modulo-$2^{16}$ phase accumulation with a 65{,}520-sample period and matching the DDC averaging window to the same length.
Hardware measurements in a purely digital loop-back configuration confirm the complete removal of the spurious tones, in agreement with the digital-twin predictions.
The proposed modification introduces a limited increase in FPGA resource utilization, which remains well within the available device resources and is acceptable given the achieved spectral improvements.

\bibliography{sn-bibliography}

\begin{thebibliography}{9}
\providecommand{\natexlab}[1]{#1}
\providecommand{\url}[1]{{#1}}
\providecommand{\urlprefix}{URL }
\providecommand{\doi}[1]{\url{https://doi.org/#1}}
\providecommand{\eprint}[2][]{\url{#2}}
 \bibcommenthead

\bibitem[{Abdkrimi et~al.(2024)Abdkrimi, Rossetto, Bourrion, Vescovi, and Hoarau}]{abdkrimi2024modeling}
Abdkrimi M, Rossetto O, Bourrion O, et~al (2024) Modeling and analysis of digital-to-analog converter non-idealities in microwave kinetic inductance detectors readout. In: 2024 IEEE 28th Workshop on Signal and Power Integrity (SPI). IEEE

\bibitem[{Abdkrimi et~al.(2025{\natexlab{a}})Abdkrimi, Rossetto, Bourrion, Vescovi, and Hoarau}]{abdkrimi2025efficient}
Abdkrimi M, Rossetto O, Bourrion O, et~al (2025{\natexlab{a}}) Efficient fpga readout architecture for mkids: A dsp-light approach. In: 2025 20th International Conference on PhD Research in Microelectronics and Electronics (PRIME), IEEE, pp 1--4

\bibitem[{Abdkrimi et~al.(2025{\natexlab{b}})Abdkrimi, Rossetto, Bourrion, Vescovi, and Hoarau}]{abdkrimi2025cordic}
Abdkrimi M, Rossetto O, Bourrion O, et~al (2025{\natexlab{b}}) Optimized fpga implementation of the cordic algorithm for a frequency multiplexed readout. In: 2025 14th Mediterranean Conference on Embedded Computing (MECO). IEEE

\bibitem[{Bounmy et~al.(2022)Bounmy, Hoarau, Mac{\'\i}as-P{\'e}rez, Beelen, Beno{\^\i}t, Bourrion, Calvo, Catalano, Fasano, Goupy et~al.}]{bounmy2022concerto}
Bounmy J, Hoarau C, Mac{\'\i}as-P{\'e}rez JF, et~al (2022) Concerto: Digital processing for finding and tuning lekids. Journal of Instrumentation 17(08):P08037

\bibitem[{Bourrion et~al.(2022)Bourrion, Hoarau, Bounmy, Tourres, Vescovi, Bouly, Ponchant, Beelen, Calvo, Catalano et~al.}]{bourrion2022concerto}
Bourrion O, Hoarau C, Bounmy J, et~al (2022) Concerto: Readout and control electronics. Journal of Instrumentation 17(10):P10047

\bibitem[{Catalano et~al.(2022)Catalano, Ade, Aravena, Barria, Beelen, Benoit, B{\'e}thermin, Bounmy, Bourrion, Bres et~al.}]{catalano2022concerto}
Catalano A, Ade P, Aravena M, et~al (2022) Concerto at apex: Installation and first phase of on-sky commissioning. In: EPJ Web of Conferences, EDP Sciences, p 00010

\bibitem[{Day et~al.(2003)Day, LeDuc, Mazin, Vayonakis, and Zmuidzinas}]{Day2003}
Day PK, LeDuc HG, Mazin BA, et~al (2003) {A broadband superconducting detector suitable for large arrays}. Nature 425(6960):817--821

\bibitem[{Klutsch(2003)}]{klutsch2003modelisation}
Klutsch I (2003) Mod{\'e}lisation des supraconducteurs et mesures. PhD thesis, Institut National Polytechnique de Grenoble-INPG

\bibitem[{Ward-Thompson et~al.(2007)Ward-Thompson, Andr{\'e}, Crutcher, and et~al.}]{ward2007protostars}
Ward-Thompson D, Andr{\'e} P, Crutcher R, et~al (2007) Protostars and planets v. University of Arizona Press

\end{thebibliography}

\end{document}